\documentclass[aps,twocolumn,superscriptaddress,longbibliography,pre]{revtex4-1}
\usepackage{amssymb} 
\usepackage{graphicx,color} 
\usepackage{bbm} 
\usepackage{multirow}
\usepackage[draft]{changes}
\usepackage{amsmath}
\usepackage{mathrsfs} 
\usepackage{hyperref}
\usepackage{xcolor}
\usepackage{graphicx}
\usepackage{cancel}
\usepackage{color} 
\usepackage{multibib}
\newcites{supp}{Supplementary References}

\usepackage{tikz}
\definecolor{darkgreen}{rgb}{0,0.6,0}
\definecolor{darkblue}{rgb}{0,0,0.6}
\definecolor{darkred}{rgb}{0.6,0,0}
\definecolor{darkpurple}{rgb}{0.5,0,0.5}

\graphicspath{{figures/}}

\hypersetup{
bookmarksopen=true,
pdftitle="",
pdfauthor="", 
pdftoolbar=false, 
pdfstartview={FitH},		
pdfmenubar=true,			
pdfhighlight=/O,			
colorlinks=true,			
urlcolor=darkblue,
citecolor=darkblue,		
linkcolor=darkblue}

\newcommand{\plaind}{\mathrm{d}}
\newcommand{\dt}{\plaind t}

\newcommand{\dint}[1]{\mathchoice{\!\plaind#1\,}{\!\plaind#1\,}{\!\plaind#1\,}{\!\plaind#1\,}}

\DeclareMathOperator{\Cov}{Cov}
\DeclareMathOperator{\Var}{Var}
\DeclareMathOperator{\E}{ \mathbb{E} }
\DeclareMathOperator{\sign}{sign}
\DeclareMathOperator{\argmin}{argmin}

\newcommand{\eq}{\text{ss}}
\newcommand{\engine}{\mathrm{engine}}
\newcommand{\ext}{\text{ext}}
\newcommand{\extint}{\text{ext+int}}

\newcommand{\ave}[1]{\left\langle#1\right\rangle}

\newcommand{\tf}{t_\text{f}}
\newcommand{\lambdaf}{\lambda_\text{f}}
\newcommand{\xvo}{x_{v_0}}

\newcommand{\vgt}{v_0^\mathrm{r}}
\newcommand{\vm}{v_0^\mathrm{m}}

\newcommand{\lambdam}{\lambda^\mathrm{m}}
\newcommand{\lambdameng}{\lambda^\mathrm{engine, m}}

\newcommand{\op}{\eq}

\newcommand{\elabel}[1]{\label{eq:#1}}
\newcommand{\eref}[1]{(\ref{eq:#1})}
\newcommand{\Eref}[1]{Eq.~(\ref{eq:#1})}

\allowdisplaybreaks

\begin{document}

\title{
Active particles in moving traps: minimum work protocols and information efficiency of work extraction
}

\author{Janik Sch\"uttler}%
\affiliation{%
DAMTP, Centre for Mathematical Sciences, University of Cambridge, Cambridge CB3 0WA, UK}%

\author{Rosalba Garcia-Millan}%
\affiliation{DAMTP, Centre for Mathematical Sciences, University of Cambridge, Cambridge CB3 0WA, UK}
\affiliation{%
Department of Mathematics, King's College London, London WC2R 2LS, United Kingdom}
\affiliation{St John's College, University of Cambridge, Cambridge CB2 1TP, UK}%

\author{Michael E.~Cates}%
\affiliation{%
DAMTP, Centre for Mathematical Sciences, University of Cambridge, Cambridge CB3 0WA, UK}%

\author{Sarah A.~M.~Loos}%
\email{sl2127@cam.ac.uk}
\affiliation{%
DAMTP, Centre for Mathematical Sciences, University of Cambridge, Cambridge CB3 0WA, UK}

\begin{abstract}
We revisit the elementary problem of moving a particle in a harmonic trap in finite time with minimal work cost, and extend it to the case of an active particle.
By comparing the Gaussian case of an Active Ornstein-Uhlenbeck particle and the non-Gaussian run-and-tumble particle, we establish general principles for thermodynamically optimal control of active matter beyond specific models.
We show that the open-loop optimal protocols, which do not incorporate system-state information, are identical to those of passive particles but result in larger work fluctuations due to activity. 
In contrast, closed-loop (or feedback) control with a single (initial) measurement changes the optimal protocol and reduces the average work relative to the open-loop control for small enough measurement errors.
Minimum work is achieved by particles with finite persistence time. 
As an application, we propose an active information engine which extracts work from self-propulsion. 
This periodic engine achieves higher information efficiency with run-and-tumble particles than with active Ornstein-Uhlenbeck particles. 
Complementing a companion paper that gives only the main results \cite{garcia2024optimal}, here we provide a full account of our theoretical calculations and simulation results. We include derivations of optimal protocols, work variance, impact of measurement uncertainty, and information-acquisition costs.
\end{abstract}

\maketitle

\setcounter{equation}{0}

\section{Introduction}

Optimal control can be a powerful tool for minimizing heat dissipation, work input or entropy generation in controllable systems, making processes thermodynamically more efficient or even newly feasible under limited energy supply \cite{schmiedl2007optimal,blaber2023optimal,engel2023optimal,whitelam2023demon,loos2023universal,zhong2024beyond,zhong2022limited}.
The thermodynamic cost is crucial in small-scale systems due to high frictional losses and fluctuations, making thermodynamic optimization particularly important.
%
It is well established that even for passive systems, optimal control can lead to nonintuitive solutions, such as nonsmooth external driving protocols~\cite{schmiedl2007optimal,blaber2021steps,engel2023optimal,whitelam2023demon,loos2023universal}.
In contrast, optimal control in active matter, characterized by self-driven particles with rich dynamics, has only recently been explored~\cite{davis2024active, loos2024smooth,casert2024learning,gupta2023efficient,cocconi2023optimal,cocconi2024efficiency,casert2024learning,baldovin2023}. 
To uncover fundamental principles of thermodynamically optimal control in active systems, characterized by memory in the form of persistence, Gaussian or non-Gaussian statistics, and non-thermal fluctuations, it is essential to investigate simplified, paradigmatic models.

In this work, we study the elementary problem of shifting a trapping potential containing a single active particle to a target position in finite time and in one dimension. The particle is immersed in a viscous fluid which generates friction and acts as a heat bath.
We optimize the dragging protocol to minimize the experimenter's work input, which is a directly measurable quantity \cite{ciliberto2017experiments,loos2023universal} that serves as a lower bound for the total energy expenditure of the controller. With a harmonic trap, the problem becomes linear and can be solved exactly. Despite its simplicity, the setup reveals nontrivial aspects of optimal control in active matter, as we show below.

For passive particles (PPs), this problem was solved in the seminal work of Ref.~\cite{schmiedl2007optimal}. Here, we extend these results to examine how intrinsic activity impacts transport protocols and energetics. To generalize beyond a specific model, we consider two widely studied models for self-propelled particles: the run-and-tumble particle (RTP), and the active Ornstein-Uhlenbeck particle (AOUP). 
Comparing AOUPs and RTPs explicitly reveals the influence of non-Gaussian statistics (only present for RTPs), which are common in active matter.

In the language of control theory, the dragging problem described above represents an \textit{open-loop} control, where the protocol is pre-determined without specific knowledge of the current system state. A more sophisticated approach is \textit{closed-loop} or feedback control, which dynamically adjusts the protocol based on real-time knowledge of the system's state, thereby allowing for more precise manipulation \cite{bechhoefer2005feedback}. 
As an analytically tractable example of a closed-loop control problem, we study how a single initial measurement of the system state affects the optimal driving protocol, generalizing previous studies on PPs \cite{abreu2011extracting,whitelam2023demon} to the active case.

We build on these results to construct a minimal, thermodynamically optimized engine that extracts work using a single measurement per cycle.
As this extraction relies on measurement information, the system can be considered  an ``information engine''~\cite{saha2021maximizing,parrondo2015thermodynamics,saha2023information}. However, since it draws work from particle activity, similar to active heat engines~\cite{krishnamurthy2016micrometre,holubec2020active,fodor2021active,speck2022efficiency,zakine2017stochastic,szamel2020single,saha2023information}, we refer to it as an ``active information engine.'' 
To evaluate the efficiency of information-to-work conversion, we calculate a fundamental limit to the thermodynamic cost associated with the required real-time measurements \cite{malgaretti2022szilard,sagawa2010generalized,kullback2013topics,cao2009thermodynamics,parrondo2015thermodynamics}.

The main results are presented in a companion paper \cite{garcia2024optimal}, which summarizes the key principles of optimal control in active matter that we have identified. In this paper, we provide a full theoretical account, detailed derivations, and additional analytical and numerical results to support our main conclusions.

We include a comprehensive discussion of non-equilibrium fluctuations and averages over initial conditions for open- and closed-loop control (Sec.~\ref{sec:Model}), along with a detailed derivation of the optimal protocols using variational calculus for both open-loop (Sec.~\ref{sec:open-loop})---for which we also include a derivation of the variance of the work fluctuations (see Sec.~\ref{sec:variance-work})---and closed-loop control (Sec.~\ref{sec:closed-loop}). 
Lastly, we present a detailed discussion of the impact of measurement uncertainty for the closed-loop control problem and the active information engine (Sec.~\ref{sec:epsilon}), a derivation of the information-acquisition costs (Sec.~\ref{sec:infoengine}), and numerical results for the work distribution and fluctuations of the engine.
We also consider an alternative cost functional, which explicitly includes a lower bound for the work done by the active particle itself in addition to the experimenter's work input (Sec.~\ref{sec:work-functionals}). 
We conclude in Sec.~\ref{sec:conclusion} and cover some further technical details in appendices.

\section{Model}\label{sec:Model}

In this section, we define the active Ornstein-Uhlenbeck particle (AOUP) and the run-and-tumble particle (RTP) models, along with the optimal control problem considered in this paper. We then introduce the technical tools required to analyze the optimal control problem, both with and without an initial measurement.

\subsection{Equations of motion and steady state}

We consider a one-dimensional model of a self-propelled particle at position $x$ in a harmonic trapping potential
\begin{equation}
  V(x,\lambda)= \frac12 k (x-\lambda)^2 \ ,
\end{equation}
where $k$ is the stiffness and $\lambda$ is the center of the trap.
The particle's motion follows the
overdamped Langevin equation
\begin{equation}
\elabel{oLE}
  \dot{x}(t) = -\frac{\partial}{\partial x} V(x,\lambda) + v(t) + \sqrt{2D} \ \xi(t) \ ,
\end{equation}
where $v$ denotes the self-propulsion, $D$ is a thermal diffusion constant, and $\xi$ is a unit-variance Gaussian white noise, $\ave{\xi(t)}=0$, $\ave{\xi(t)\xi(t')}= \delta(t-t')$.
Here, $\ave{\bullet}$ denotes the average over many noise realizations.
We have absorbed the friction constant and $k_\mathrm{B}$ into other parameters.
We consider two standard models of active motility: 
(i) an AOUP, where
\begin{subequations}\label{eq:v}
\begin{equation}\label{eq:v_AOUP}
  \tau \dot{v}(t) = - v(t) +  \sqrt{2D_v} \ \xi_v(t) \ ,
\end{equation}
where $D_v$ represents the self-propulsion ``diffusion'' constant, and $\xi_v$ is a zero-mean, unit-variance Gaussian white noise independent of $\xi$;
and (ii) an RTP, 
where the self-propulsion
is a dichotomous, or telegraphic, Poissonian noise
that randomly reassigns values 
\begin{equation}
  v = \pm  \omega
\end{equation}
\end{subequations}
with a rate of $1/(2\tau)$, which corresponds to a tumbling rate of $1/\tau$ \cite{schnitzer1990strategies, schnitzer1993theory, schneider2019optimal, cates2012diffusive, garcia2021run}.
A direct comparison between the two active models can be established up to second-order by equating the variance of active fluctuations (derived in  App.~\ref{sec:covariance})
\begin{equation}\label{eq:rtp_aoup_match}
  D_v/\tau \equiv {\omega}^2 
\end{equation}
which we take to hold throughout this paper so as to enable direct comparisons between the two cases.
Comparing AOUPs and RTPs enables us to isolate the effect of introducing non-Gaussianity in the form of a Poissonian noise (RTPs) \cite{lee2022effects}. We compare both active particle models against the case of a PP, where $v(t)\equiv 0$.
Both active models reduce to PPs under the limits $\omega\to0$ or $\tau\to0$. In the limit $\tau\to\infty$, the active models yield a ballistic (undamped) motion with a constant velocity randomly set by the initial condition.

At time $t=0$, and with $\lambda = \lambda_0=0$ initially, we consider the system in a steady state which we denote by $P_\eq(x,v)$ and the average with respect to $P_\eq$ by $\ave{\bullet}$. 
For AOUPs, the joint steady state is Gaussian, and it is non-Gaussian for RTPs \cite{garcia2021run}.
The first moments in the steady state are both zero,
\begin{subequations}
\label{eq:steadystate}
\begin{align}
  \ave{x} &= 0  \ , \\
  \ave{v} &= 0  \ ,
\end{align}
and the second moments are given by
\begin{align}
\begin{split}
  \ave{v^2} &=  \omega^2 \ ,
\end{split}
  \\
\begin{split}
  \ave{x^2} &= \frac1k\left(D + \frac{\omega^2}{k + 1/\tau}\right) \ ,
\end{split}
  \\
  \ave{xv} &= \frac{\omega^2}{k + 1/\tau} \ .
\end{align}
\end{subequations}
For AOUPs, these first two moments fully characterize the steady-state distribution.
Finally, we define the conditional density as $P_\eq(x | v) = P_\eq(x, v ) / P_\eq(v)$.

\subsection{Central question: Optimization of the  average work}

Our objective is to move the trap from the origin, $\lambda_0 = 0$, at time $t = 0$, to the target position $\lambda(\tf) = \lambdaf$ within a finite time $\tf$. Due to the frictional resistance of the surrounding fluid and changes in the particle’s potential energy, this process requires external work input $W$ from the controller. 
Moving the trap by an infinitesimal distance $\plaind{\lambda}$ requires work equal to the infinitesimal change in the particle’s potential energy.
\cite{jarzynski1997nonequilibrium, crooks1998nonequilibrium}
\begin{equation}
  \delta W = \frac{\partial V(x,\lambda)}{\partial \lambda} \plaind{\lambda}
    = k (\lambda - x) \plaind{\lambda} \ ,
\end{equation}
where we use $\delta$ to denote inexact differentials. 
As a result, the total external work required to move the trap from $\lambda(0) = 0$ to $\lambda(\tf) = \lambdaf$ is
\begin{equation}
\label{eq:def_work1}
  W[\lambda(t), x(t)] := \int_0^{\tf}\plaind{t}\ \dot{\lambda} k (\lambda - x) \ .
\end{equation} 

We optimize the time-dependent protocol, denoted by $\lambda(t)$, to minimize the required \textit{average} work input
\begin{equation}\label{eq:optimalcontrol}
  \underset{\lambda(t)}{\argmin} \ave{ W }
\end{equation}
with boundary condition
\begin{subequations}\label{eq:optimalcontrol_bc}
\begin{align}
  \lambda(0) &= 0 \ ,
  \\
  \lambda(\tf) &= \lambdaf \ .
\end{align}
\end{subequations}
We consider different ensemble averages $\ave{\bullet}$, defined in the following subsection.

\subsection{Hierarchy of averages}\label{app:ensembles}

We aim to calculate the optimal protocol $\lambda(t)$ in two distinct scenarios: i) without any additional information about the state of $x$ and $v$ (open-loop case), and ii) with knowledge of the state of $x$ or $v$ at $t=0$ obtained through an initial measurement (closed-loop case). 
In this subsection, we set the technical tools to analyze these two scenarios. 
For now, we assume that the measurement is exact, and defer the discussion of measurement uncertainty to Sec.~\ref{sec:epsilon}. 

We first consider the case without additional information about $x$ and $v$. In this case, the system is initially (at $t=0$) found in a steady state. We define the \emph{stationary ensemble} as the ensemble consisting of all trajectories $\{x(t),v(t)\}_{0\leq t \leq \tf}$ of the stochastic process described by Eqs.~\eqref{eq:oLE} and \eqref{eq:v}, where $x(0)$ and $v(0)$ are random variables drawn from the stationary distribution.
The corresponding average over the noises $\eta$ and $\xi$ in the stationary ensemble is given by $\ave{\bullet}_{\op}$. 
The stationary ensemble has the following average initial conditions
\begin{subequations}\label{eq:ic_stationary}
\begin{align}
  \ave{x(0)}_\op &= 0 \ ,
  \\
  \ave{v(0)}_\op &= 0 \ .
\end{align}
\end{subequations}
Note that the stationary ensemble consists of trajectories which are initially in a steady state at $t=0$. However, once the controller applies the protocol at $t>0$, these trajectories are generally not stationary at intermediate times $t > 0$. 

In contrast, when the initial state of the system is known to be  $x(0) = x_0$ and $v(0) = v_0$, we define the \emph{conditional ensemble}. 
This ensemble consists of all trajectories $\{x(t),v(t)\}_{0\leq t \leq \tf}$ of the stochastic process Eqs.~\eqref{eq:oLE} and \eqref{eq:v}, with the initial state of all trajectories is fixed to $x(0)=x_0$ and $v(0)=v_0$. The corresponding average is the conditional average $\ave{\bullet | x(0) = x_0, v(0) = v_0}_\op$. 
To abbreviate notation, we define the following shorthand 
\begin{equation}
  \ave{\bullet }_{x_0,v_0} := \ave{\bullet | x(0) = x_0, v(0) = v_0}_\op \ .
\end{equation}
Accordingly, the initial condition in the conditional ensemble is
\begin{subequations}\label{eq:ic_conditional}
\begin{align}
  \ave{x(0)}_{x_0,v_0} &= x_0 \ ,
  \\
  \ave{v(0)}_{x_0,v_0} &= v_0 \ .
\end{align}
\end{subequations}
Because the system is stationary at $t=0$, $x_0$ and $v_0$ are samples from the stationary state. 
As a result, the conditional ensemble is related to the stationary ensemble through the law of total expectation
\begin{equation}\label{eq:tower_property}
    \ave{\bullet}_\op = \E_{x_0,v_0}[ \ave{\bullet}_{x_0,v_0} ] 
    \ ,
\end{equation}
where $\E_{x_0,v_0}$ is the steady average with respect to $(x_0,v_0)$, defined as
\begin{equation}
    \E_{x_0,v_0}[\bullet] := \iint_{\mathbb{R}} \plaind{x_0}\plaind{v_0} \, \bullet \ P_\mathrm{ss}(x_0, v_0) \ .
\end{equation}
For RTPs, the integral over $v_0$ reduces to a sum over $\pm \omega$.

In the remainder of the paper, we will show that most of the novel physics of the closed-loop control derives from measurements of the self-propulsion rather than of the positional measurements. In order to analyze closed-loop control with only self-propulsion measurements, we define the \emph{partial conditional ensemble}, consisting of all trajectories $\{x(t),v(t)\}_{0\leq t \leq \tf}$, where the initial self-propulsion is fixed to $v(0) = v_0$. 
The corresponding average is $\ave{\bullet | v(0) = v_0}$, and introduce the shorthand notation
\begin{equation}
  \ave{\bullet}_{v_0} := \ave{ \bullet | v(0) = v_0}_\op \ .
\end{equation}
The law of total expectation establishes the following relation between averages,
\begin{subequations}\label{eq:tower_property2}
\begin{align}
  \ave{\bullet} &= \E_{v_0}[ \ave{\bullet}_{v_0} ] \ ,
  \\
  \label{eq:tower_property_partialcondens}
  \ave{\bullet }_{v_0} &= \E_{x_0}[\ave{\bullet }_{v_0,x_0} \vert v_0 ] \ ,
\end{align}
\end{subequations}
where the steady state conditional average is defined using $P_\eq(x|v)$ as
\begin{equation}
  \E_{x_0}[\bullet | v_0] := \int_{\mathbb{R}} \plaind{x_0} \, \bullet \ P_\mathrm{ss}(x_0| v_0) \ .
\end{equation}
Because $x$ and $v$ are correlated, measuring $v_0$ reveals information about $x(0)$. 
In particular, given a measurement $v_0$, the average initial position of the particle is
\begin{align}\label{eq:x_given_v_ss}
\begin{split}
  \xvo &:= \E_{x_0}[\ave{x(0)}_{x_0,v_0} | v_0] 
  \\
  &\;= \frac{ \E_{x,v}[xv] }{ \E_{x,v}[v^2] } \, v_0
  = \frac{v_0}{k + 1/\tau} \ .
\end{split}
\end{align}
Together, the initial conditions in the partial conditional ensemble are
\begin{subequations}\label{eq:ic_pce}
\begin{align}
  \ave{x(0)}_{v_0} &= \xvo  \ ,
  \\
  \ave{v(0)}_{v_0} &= v_0 \ .
\end{align}
\end{subequations}

The ensemble with a measurement of the initial position only (with conditional average $\ave{\bullet}_{x_0}$) can be defined analogously.

In the derivations that follow, many steps are identical across the different ensembles. When this is the case, we will use the placeholder notation $\ave{\bullet}_\alpha$ to represent a general average.

\subsection{Averages of position and self-propulsion}
\label{sec:averages}

We now derive expressions for the average position and self-propulsion, which are required for evaluating the average work. 
Averaging Eqs.~\eqref{eq:oLE} and \eqref{eq:v} across any of the three ensembles (stationary, conditional, or partial conditional) leads to the following differential equation, identical in form for RTP and AOUP models,
\begin{subequations}\label{eq:ave_le}
\begin{align}
    \frac{\plaind}{\plaind t} \ave{{x}(t)}_\alpha &= k(\lambda - \ave{x(t)}_\alpha) + \ave{v(t)}_\alpha \ ,
    \\
    \frac{\plaind}{\plaind t} \ave{v(t)}_\alpha &= -\frac1{\tau} \ave{v(t)}_\alpha \ ,
\end{align}
\end{subequations}
where $\alpha$ denotes the appropriate ensemble.
Solving Eq.~\eqref{eq:ave_le} with the initial conditions in Eqs.~\eqref{eq:ic_stationary}, \eqref{eq:ic_conditional} and \eqref{eq:ic_pce}, we find the average self-propulsion
\begin{subequations}
\label{eq:ave_selfprop}
\begin{align}
  \label{eq:ave_selfprop_conditional}
  \ave{v(t)}_{x_0,v_0} &= \ave{v(t)}_{v_0} = v_0 e^{-t/\tau} \ ,
  \\
  \label{eq:ave_selfprop_stationary}
  \ave{v(t)}_\op &= 0 \ .
\end{align}
\end{subequations}
Next, we solve the average particle position and obtain
\begin{subequations}\label{eq:ave_x_formal1}
\begin{align}
  \ave{x(t)}_\op[\lambda(t)] &= 
  k \int_{0}^t\dint{t'} \lambda(t')e^{-k(t-t')} \ ,
  \\
\nonumber
  \ave{x(t)}_{x_0,v_0}[\lambda(t)] &= 
  x_0e^{-kt} + \frac{v_0}{k-1/\tau} \left(e^{-t/\tau} - e^{-kt}\right) 
  \\
  &+ k \int_{0}^t\dint{t'} \lambda(t')e^{-k(t-t')} \ ,
  \\
\nonumber
  \ave{x(t)}_{v_0}[\lambda(t)] &= 
  \frac{v_0 e^{-kt}}{k + 1/\tau} +\frac{v_0}{k-1/\tau} \left(e^{-t/\tau} - e^{-kt}\right) 
  \\
  &+ k \int_{0}^t\dint{t'} \lambda(t')e^{-k(t-t')} \ .
\end{align}
\end{subequations}
These equations highlight the dependence of $\ave{x}_\alpha$ on the protocol $\lambda$.
For brevity, we typically omit the explicit dependence on $\lambda(t)$ and abbreviate the average particle position as $\ave{x(t)}_\alpha$. 
By rewriting Eq.~\eqref{eq:ave_le}, the protocol can be expressed as a function of the average particle position
\begin{equation}
\label{eq:lambda_from_x}
  \lambda_\alpha(t) = \frac{\ave{\dot{x}(t)}_\alpha - \ave{v(t)}_\alpha}{k} + \ave{x(t)}_\alpha \ ,
\end{equation} 
for intermediate times $0 < t < \tf$.

In addition to the first moments discussed in this subsection, we also require the second moments of these quantities at a few specific points later in the text. For completeness, their derivation is provided in App.~\ref{sec:covariance}.
Importantly, all first and second moments of $x$ and $v$ are identical for RTPs and AOUPs upon identifying Eq.~\eqref{eq:rtp_aoup_match} (explicitly discussed in App.~\ref{sec:covariance}). For this reason many--but not all--results we derive in this paper are equivalent for RTPs and AOUPs.

\subsection{Average work}

We now discuss averages of the work. Taking the ensemble average of Eq.~\eqref{eq:def_work1}, we find the average work
\begin{align}
\label{eq:workfunctional}
\begin{split}
  \ave{W}_\alpha 
  &= \int_0^{\tf}\plaind{t}\ \dot{\lambda}(t) k \left(\lambda(t) - \ave{x(t)}_\alpha \right)  
  \\
  &= W \left[ \lambda(t), \ave{x(t)}_\alpha \right]
  \ .
\end{split}
\end{align}
Substituting $\lambda = \lambda_\alpha$ from Eq.~\eqref{eq:lambda_from_x} into $W[\lambda, x]$, we define
\begin{align}\label{eq:W2}
  W[\ave{x}_\alpha] 
  &:= W  \left[  \frac{\ave{\dot{x}(t)}_\alpha - \ave{v(t)}_\alpha}{k} + \ave{x(t)}_\alpha, \ave{x(t)}_\alpha \right] \ ,
\end{align}
which depends on $\ave{x}_\alpha$ but not explicitly on $\lambda$.
Note that $W[\ave{x}_\alpha]$ also depends on $\ave{v}_\alpha$, but we omit this dependence in the notation as $\ave{v}_\alpha$ does not depend on $\ave{x}_\alpha$ and thus does not affect the optimization of $W[\ave{x}_\alpha]$.
Furthermore, note that the functional in Eq.~\eqref{eq:W2} is linear in the trajectory, meaning $\ave{W[x(t)]}_{\alpha} = W[\ave{x}_\alpha]$. 
Using partial integration in Eqs.~\eqref{eq:W2} and \eqref{eq:workfunctional}, we find the explicit expression 
\begin{equation}
\label{eq:work_ext_functional}
\begin{split}
  \ave{W}_\alpha &= W[\ave{x}_\alpha]
  = \frac{k}{2}\left[(\lambda(t) - \ave{x(t)}_\alpha)^2\right]_0^{\tf} 
  \\
  &\quad + \int_0^{\tf}\plaind{t}\ 
  \left( \ave{\dot{x}(t)}_\alpha^2 - \ave{\dot{x}(t)}_\alpha \ave{v(t)}_\alpha \right)\,.
\end{split}
\end{equation}
Note that, despite the non-linear form of Eq.~\eqref{eq:work_ext_functional}, $W[\ave{x}_\alpha]$ remains linear in the trajectory and the law of total expectation holds, $\ave{W[x(t)]}_{\alpha} = W[\ave{x}_\alpha]$, as inherited from Eq.~\eqref{eq:workfunctional}.
We use $W[\ave{x}_\alpha]$ as the main cost functional in the optimization. 
The optimization problem defined in Eq.~\eqref{eq:optimalcontrol} can now be formally expressed as
\begin{align}\label{eq:optimaltransport_work_ext}
  \argmin_{\ave{x(t)}_\alpha} W [ \ave{x(t)}_\alpha ] \ ,
\end{align}
with initial conditions for $\ave{x(0)}_\alpha$ and $\ave{v(0)}_\alpha$ provided by Eqs.~\eqref{eq:ic_stationary}, \eqref{eq:ic_conditional}, and \eqref{eq:ic_pce}. 
Once the optimal $\ave{x}_\alpha$ is found, the corresponding optimal protocol $\lambda$ can be obtained using Eq.~\eqref{eq:lambda_from_x}, while ensuring the boundary conditions in Eq.~\eqref{eq:optimalcontrol_bc} are satisfied. 

Finally, Eq.~\eqref{eq:work_ext_functional} shows that the functional $\ave{W }_\alpha$ depends only on first moments of position and self-propulsion. Since these moments are identical for both RTPs and AOUPs (see Secs.~\ref{sec:averages} and \ref{sec:covariance}), we conclude that the minimizers $\ave{x(t)}_\alpha$ and the corresponding optimal protocols are \textit{identical} for both models, as well as for any other model obeying Eq.~\eqref{eq:ave_le} on average.

\section{Optimal open-loop control}
\label{sec:open-loop}

In this section, we discuss the optimal transport problem in Eq.~\eqref{eq:optimaltransport_work_ext} for the stationary ensemble, i.e., the minimization of the work Eq.~\eqref{eq:work_ext_functional} in the open-loop control case, where no measurements are taken. 

\subsection{Optimal protocol}

In the stationary ensemble, the average self-propulsion vanishes (see Eq.~\eqref{eq:ave_selfprop_stationary}). 
As a result, the average external work input in the open-loop control reduces to
\begin{equation}\label{eq:work-functional-ext-open-loop}
  \ave{W}_\op = \frac{k}{2}\left[(\lambda - \ave{x}_\op)^2\right]_0^{\tf} + \int_0^{\tf}\plaind{t}\ 
  \ave{\dot{x}}^2_\op \ .
\end{equation}
Since $\ave{x}_\op$ is independent of the activity [as shown in Eq.~\eqref{eq:ave_x_formal1}], the optimization of $\ave{W}_\op$ is equivalent to that of a passive particle, as discussed in Ref.~\cite{schmiedl2007efficiency}.
Thus, the optimal particle position and protocol are identical to those in the passive case, and given by
\begin{subequations}\label{eq:opt_stationary}
\begin{align}
    \ave{x(t)}_\op &=  \frac{\lambdaf}{\tf + 2/k} t  
    \ ,
    \\
    \lambda_\op(t) &= \frac{\lambdaf}{\tf + 2/k}\left(t+\frac{1}{k}\right)
  \ , \qquad\text{for } 0 < t < \tf,
\end{align}
\end{subequations}
with boundary condition $\lambda_\op(0) = 0, \lambda_\op(\tf) = \lambdaf$.
A notable feature of the protocol $\lambda$ are the two symmetric discrete jumps at $t=0$ and $t=\tf$, with lengths
\begin{equation}
  \lambda_\op(0^+) - \lambda_\op(0) = \lambda_\op(\tf) - \lambda_\op(\tf^-) = \frac{\lambdaf}{k\tf + 2} \ .
\end{equation}
These jumps are connected by a linear dragging regime for $0 < t < \tf$. 

Moreover, the average work required to execute the optimal protocol for an active particle is the equivalent to that of a passive particle
\begin{equation}\label{eq:opt_work_ext_openloop}
  \ave{W}_\op = \frac{\lambdaf^2}{\tf + 2/k}  \ .
\end{equation}
Thus, in the case of open-loop control, activity does not influence the optimal protocol or the average work input.

\subsection{Variance of work} \label{sec:variance-work}

\begin{figure}
  \centering
  \includegraphics{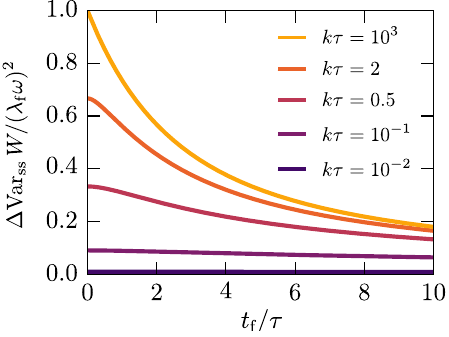}
  \caption{
  Difference in work fluctuations $\Delta \Var_\op W := \Var_\op W - \Var_\op W^{\text{passive}}$ between active and passive systems during the open-loop optimal protocol $\lambda_\op$, Eq.~\eqref{eq:DeltaVar}, as a function of dimensionless parameters. Fluctuations of the work in the active system are greater than in the passive system. Equality is reached only in the quasi-static limit, where the fluctuations vanish in both cases.
  }
  \label{fig:dVarW}
\end{figure}

To quantify the work fluctuations, we calculate the variance
\begin{align}
  \Var_\op W = \ave{ W^2 }_\op - \ave{ W }^2_\op
\end{align}
for the optimal protocol in Eq.~\eqref{eq:opt_stationary}. 
Using the definition in Eq.~\eqref{eq:def_work1}, the variance can be written as
\begin{equation}
\label{eq:var_W}
  \Var_\op W = k^2 \iint_0^{\tf}\dt_1 \dt_2\, \dot{\lambda}(t_1) \dot{\lambda}(t_2)  \, \Cov_\op[ x(t_1) x(t_2)] \ .
\end{equation}
The covariance in the stationary ensemble, given by Eq.~\eqref{eq:Covx_1}, features two terms. 
First, we evaluate the variance arising from the passive terms in $\Cov_\op [x(t) x(t')]$ proportional to $D$, leading to
\begin{align}
   \Var_\op W^{\rm passive} := \Var_\op W\vert_{\omega^2 = 0} &=   \frac{ 2D \lambdaf^2 }{ \tf + 2/k} \ .
    \label{eq:VarW_passive_nomeasurement}
\end{align}
Next, we evaluate the variance arising from the terms in $\Cov_\op [x(t) x(t')]$ proportional to $\omega^2$, and find that the correction is identical for both RTPs and AOUPs: 
\begin{align}\label{eq:DeltaVar}
  \Var_\op W  - \Var_\op W^{ \text{passive}} = \lambdaf^2  \omega^2 \Upsilon(\tf/\tau, k\tau)\geq 0 \ ,
\end{align}
where
\begin{equation}
\begin{split}
  &\Upsilon(\tf/\tau, k\tau) 
  \\
  &= \frac{ 2 }{ \left( \frac{\tf}{\tau} + \frac{2}{k\tau} \right)^2 }
    \left(
      \frac{\tf}{\tau} + \frac{\frac{2}{k\tau}+(1-k\tau)(1-e^{-\tf/\tau})}{1 + k\tau} 
    \right)
\end{split}
\end{equation}
is plotted in Fig.~\ref{fig:dVarW}.
Thus, the variance of the work is always increased by the activity, consistent with the concept of an increased effective temperature. This additional variance  vanishes only in the trivial limits:  $\lambdaf\to 0$; $\omega \to 0$ (where  active particles become passive), and in the quasistatic limit $\tf\to \infty$, where $\Var_\op W$ vanishes for all models. 

It follows that activity does not provide an advantage in the open-loop case: while the minimum average work and optimal protocol remain identical to the passive case, active systems exhibit higher work uncertainty compared to passive systems.
Furthermore, at the level of first and second moments, the (non-)Gaussianity is irrelevant for the open-loop control. However, differences arise in higher moments---for example, while the positional distribution remains Gaussian for PPs and AOUPs throughout the protocol, the non-Gaussian distribution of RTPs is generally modified by the dragging process \cite{garcia2021run}.


\section{Optimal closed-loop control}
\label{sec:closed-loop}

We now turn to the case where a feedback controller takes an initial real-time measurement before deciding and executing the protocol. 
From a thermodynamic perspective, a closed-loop controller that measures noisy particle states and adjusts the potential to achieve a target based on the measurement outcomes, could be considered a mesoscopic version of a ``Maxwell demon.'' According to Landauer's principle, the measurements required for closed-loop control are associated with thermodynamic costs~\cite{parrondo2015thermodynamics,jun2014high,berut2012experimental,ribezzi2019large}, which we discuss in Sec.~\ref{sec:epsilon}. 

The optimal closed-loop protocol can be obtained by the same optimization procedure as before, but now incorporating the measured states $x(0)=x_0,v(0)=v_0$ as initial condition. 
We consider two scenarios: one where the controller takes dual measurements of both position and self-propulsion (using the conditional ensemble defined in Sec.~\ref{app:ensembles}), and another where only the self-propulsion is measured, without a position measurement (using the partial conditional ensemble).
For now, we assume exact measurements, and we defer the discussion of measurement uncertainty to Sec.~\ref{sec:epsilon}.

\subsection{Closed-loop control with measurements of position and self-propulsion}

\begin{figure*}
\includegraphics[width=\textwidth]{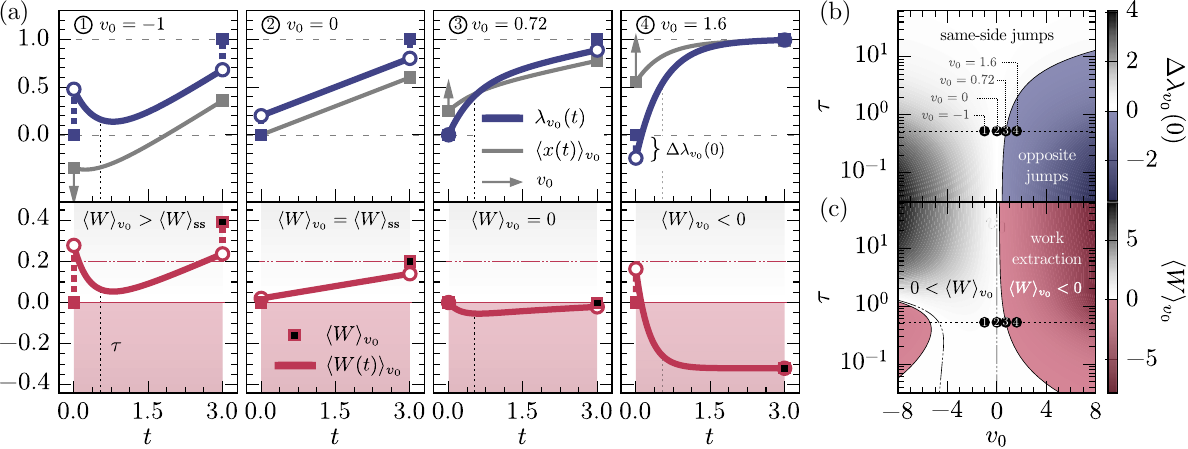}
\caption{
Optimal transport protocol for the closed-loop control problem.
(a) Top: Protocol $\lambda_{v_0}(t)$ (blue, Eq.\eqref{eq:lambdaoptactive}), average particle position $\ave{x(t)}_{v_0}$ (grey, Eq.~\eqref{eq:xaveopt1}), and self-propulsion measurement $v_0$ (arrows).
Bottom: Cumulative average work $\ave{W(t)}_{v_0}$ (red line, Eq.~\eqref{eq:cumulative_work}) and total work per protocol execution $\ave{W}_{v_0}$ (red/black square, Eq.~\eqref{eq:avg_work_partialconditional}). Red shade indicates $\ave{W(t)} \leq 0$.
(b) Initial protocol jump $\Delta \lambda_{v_0}(0)$ (Eq.~\eqref{eq:initial_jump}), with blue regions indicating initial jumps opposite the trap target, $\sign (\Delta \lambda_{v_0}(0)) \neq \sign (\lambdaf) $.
(c) Work per measurement $\langle W \rangle_{v_0}$ (Eq.~\eqref{eq:avg_work_partialconditional}), with the red region indicating work extraction, $\langle W \rangle_{v_0} < 0$.
The fixed parameters for all plots are $\tau = 0.525$, $\lambdaf = 1$, $k = 1$, and $\tf = 3$, unless varied.
}
\label{fig:work_incs}
\end{figure*}

We now discuss the optimal transport problem from Eq.~\eqref{eq:optimaltransport_work_ext} in the conditional ensemble. The corresponding cost functional in the conditional ensemble can be obtained from Eq.~\eqref{eq:work_ext_functional} as
\begin{equation}
\begin{split}
  \ave{W}_{x_0,v_0} &= \frac{k}{2}\left[(\lambda - \ave{x}_{x_0,v_0})^2\right]_0^{\tf} 
  \\
  & \ + \int_0^{\tf}\plaind{t}\ 
  \left( \ave{\dot{x}}_{x_0,v_0}^2 - \ave{\dot{x}}_{x_0,v_0} \ave{v}_{x_0,v_0} \right)
  \ .
\end{split}
\end{equation}
We begin by optimizing the bulk term 
\begin{equation}
  \argmin_{\ave{x(t)}_{x_0,v_0}} \int_0^{\tf}\plaind{t}\ 
     \left( \ave{\dot{x}}_{x_0,v_0}^2 - \ave{\dot{x}}_{x_0,v_0} \ave{v}_{x_0,v_0} \right) \ ,
\end{equation}
with respect to $\ave{x}_{x_0,v_0}$ which leads to the following Euler-Lagrange equation
\begin{equation}\label{eq:eulerlagrange}
    \ave{\ddot{x}}_{x_0,v_0} = \frac12 \ave{\dot{v}}_{x_0,v_0} \ .
\end{equation}
By integrating Eq.~\eqref{eq:eulerlagrange} and using the solution for $\ave{v(t)}_{x_0,v_0}$ in Eq.~\eqref{eq:ave_selfprop_conditional}, along with the initial condition $\ave{x(0)}_{x_0,v_0} = x_0$, we obtain the solution
\begin{align}\label{eq:opt_avg_x_raw}
  \ave{x(t)}_{x_0,v_0} &= x_0  + \mathcal{C} t  + \frac{\tau v_0}{2} (1-e^{-t/\tau}) \ , 
\end{align}    
in terms of an unknown constant driving velocity $\mathcal{C}$. Evaluating $\ave{W}_{x_0,v_0}$ using the solution \eqref{eq:opt_avg_x_raw} and optimizing the resulting expression with respect to $\mathcal{C}$ yields
\begin{equation}
\elabel{opt_m_general}
  \mathcal{C} = \frac{ \lambdaf - x_0 - \frac12 \tau v_0 (1 -e^{-\tf/\tau}) }{\tf+2/k} \ .
\end{equation}
Substituting $\mathcal{C}$ into the solution in Eq.~\eqref{eq:opt_avg_x_raw}, we find the optimal average position and corresponding optimal protocol
\begin{subequations}\label{eq:opt_conditional}
\begin{align}
  \label{eq:opt_x_conditional}
  \ave{x(t)}_{x_0,v_0} &= x_0
  + \frac{d_{x_0,v_0}}{\tf + 2/k} t  + \frac{\tau v_0}{2}  (1-e^{-t/\tau})  \ ,
  \\
  \label{eq:opt_lambda_conditional}
  \lambda_{x_0,v_0}(t) &= \ave{x(t)}_{x_0,v_0} + \frac{d_{x_0,v_0}}{k\tf + 2} - \frac{v_0 }{2 k} e^{-t/\tau},
\end{align}
with boundary conditions $\lambda_{x_0,v_0}(0) = 0, \lambda_{x_0,v_0}(\tf) = \lambdaf$. The quantity $d_{x_0,v_0}$ is an effective distance, defined as
\begin{equation}
\elabel{d_eff}
  d_{x_0,v_0} = \lambdaf - x_0 - \frac{\tau v_0}{2}  \left(1-e^{-\tf/\tau}\right)
  \ ,
\end{equation}
\end{subequations}
which corresponds to the distance between the particle's initial position and the target position of the trap (first two terms) minus the distance covered ``for free'' by the particle's self-propulsion before its orientation decorrelates.

The optimal value of the average work input is given by
\begin{align}\label{eq:avg_work_conditional}
    \ave{W}_{x_0,v_0}  &=  
    -\frac{k x_0^2}2 
    - {\frac{\tau v_0^2}{8}} \left(1- e^{-2\tf/\tau} \right) 
    + \frac{d^2_{x_0,v_0}}{\tf + 2/k} \ ,
\end{align}
which shows that the measurements contribute to work extraction  through two different physical mechanisms. 
The term $- \frac12 k x_0$ represents the total potential energy stored in the initial configuration that the optimal protocol extracts. This term also appears in closed-loop control of PPs \cite{abreu2011extracting}. Furthermore, there is an activity-dependent contribution that is non-positive and proportional to $v_0^2$ which arises from work extraction due to the activity \cite{malgaretti2022szilard}.
Unlike work extraction from the potential energy, enabled by measuring $x_0$, extracting work from activity requires a finite protocol duration, $\tf > 0$. We provide a more detailed discussion of this mechanism in Sec.~\ref{sec:workpm}. 

The physics of the $x$-measurement in the active particles considered here is identical to that of PPs, as discussed in Ref.~\cite{abreu2011extracting}. 
Since $x$-measurements on active particles do not entail new physical phenomena, we will focus on the problem of $v$-measurements in the following.

\subsection{Closed-loop control with only self-propulsion measurement}

\subsubsection{Optimal protocol}

The optimal protocol can be derived from the previous results by using the relationship between the conditional and partial conditional ensembles (see Sec. \ref{app:ensembles}).
Specifically, a measurement of $v$ provides information about the average initial position $\ave{x(0)}_{v_0} = \xvo = v_0/(k+1/\tau)$, as described by Eq.~\eqref{eq:x_given_v_ss}.
Together with the law of total expectation Eq.~\eqref{eq:tower_property_partialcondens}, we obtain both the optimal protocol and optimal average particle trajectory
\begin{subequations}\label{eq:opt_partialconditional}
\begin{align}
\begin{split}
\label{eq:xaveopt1}
  \ave{x(t)}_{v_0} &= \E_{x_0}[\ave{x(t)}_{x_0,v_0} | v_0] 
  \\
  &= \xvo
  + \frac{d_{v_0}}{\tf + 2/k} t  + \frac{\tau v_0}{2}  (1-e^{-t/\tau})  \ ,
\end{split}
  \\
  \label{eq:lambdaoptactive}
  \lambda_{v_0}(t) &= \ave{x(t))}_{v_0} + \frac{d_{v_0}}{k\tf + 2} - \frac{v_0}{2 k} e^{-t/\tau}
  \ ,
  \\
\begin{split}
   d_{v_0} &= \E_{x_0}[d_{x_0,v_0} | v_0] 
   \\
   &= \lambdaf - \xvo  - \frac{\tau v_0}{2}  \left(1-e^{-\tf/\tau}\right) \ ,
\end{split}
\end{align}
\end{subequations}
with boundary condition $\lambda_{v_0}(0) = 0, \lambda_{v_0}(\tf) = \lambdaf$, as before. 
When $v_0 = 0$, the protocol $\lambda_{v_0}(t)$ simplifies to the open-loop protocol given by Eqs.~\eqref{eq:opt_stationary}, which is characterized by a linear dragging regime with symmetric jumps. This case is illustrated in the second example of Fig.~\ref{fig:work_incs}(a).
The remaining panels in Fig.~\ref{fig:work_incs}(a) show additional examples of the protocol $\lambda_{v_0}(t)$ for  non-zero self-propulsion values. In contrast to the $v_0 = 0$ case, these protocols are non-linear at $0 < t < \tf$. The non-linear behavior dominates for times up to the persistence time, $t \ll \tau$, while the linear parts take over for $t \gg \tau$.
As with the open-loop case, the protocol $\lambda_{v_0}$ remains discontinuous at the initial and final times, $t=0$ and $t=\tf$, with discontinuities
\begin{subequations}
\begin{align}
\begin{split}
  \label{eq:initial_jump}
  \Delta\lambda_{v_0}(0) &= \lambda_{v_0}(0^+) - \lambda_{v_0}(0) 
  \\
  &= 
  \xvo + \frac{d_{v_0}}{k\tf + 2}  - \frac{v_0}{2k}  \ ,
\end{split}
  \\
\begin{split}
\label{eq:initial_jump}
  \Delta \lambda_{v_0}(\tf) &= \lambda_{v_0}(\tf) - \lambda_{v_0}(\tf^-) 
  \\
  &= \frac{d_{v_0}}{k\tf + 2} + \frac{v_0}{2k}e^{- \tf /\tau} \ .
\end{split}
\end{align}
\end{subequations}
However, the jumps are now asymmetric, i.e., $\Delta\lambda_{v_0}(0) \neq \Delta \lambda_{v_0}(\tf)$. 
Furthermore, these jumps can be in the direction of the target ($\Delta \lambda_{v_0}(0)$ and $\lambdaf$ have the same sign), oppose the target, or be zero.
Examples for aligning, opposing, and zero jumps are shown in the initial jumps of examples 1, 4, and 3 in Fig.~\ref{fig:work_incs}(b), respectively.

\begin{figure*}
  \includegraphics[width=0.9\linewidth]{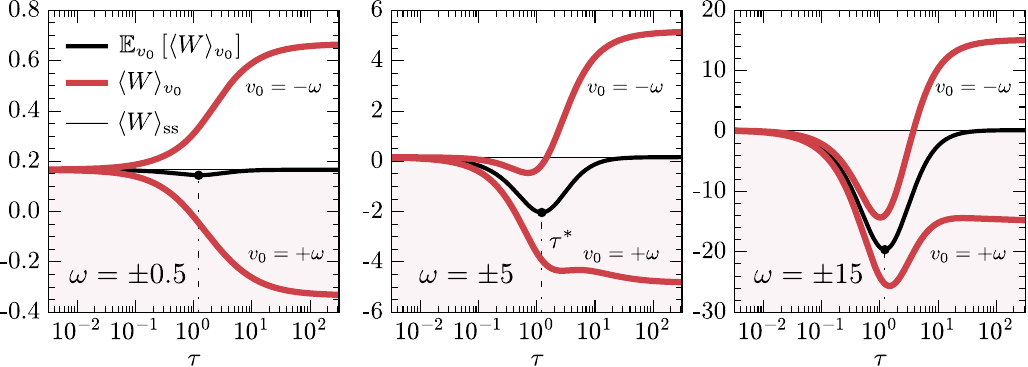}
  \caption{Work per measurement $\ave{W}_{v_0}$, Eq.~\eqref{eq:avg_work_partialconditional}, as a function of $\tau$ for a situation in which $v_0 = \pm \omega$ takes only two values of equal magnitude (e.g., in a RTP), and $k=1,\tf=4,\lambdaf=1$. Thick black line shows the average $\E_{v_0}[ \ave{W}_{v_0} ]$, Eq.~\eqref{eq:avgavg_work_partialconditional}, and thin black line shows the $\tau$-independent open-loop work $\ave{W}_\op$, Eq.~\eqref{eq:opt_work_ext_openloop}. Vertical dash-dotted line marks the optimal persistence time $\tau^*$ that minimizes $\E_{v_0}\![\ave{W}_{v_0}]$, see Eq.~\eqref{eq:tau_min}.
  Red shading marks values less than $\ave{W}_\op$.
  }
  \label{fig:Wv0_tau}
\end{figure*}

\subsubsection{Work per measurement}
\label{sec:workpm}

The initial jump is energetically costly in general, but it sets up the protocol to enable transient work extraction during the drive.
This is illustrated in the bottom panels of Fig.~\ref{fig:work_incs}(a), which shows the cumulative work up to time $t \leq \tf$ 
\begin{equation}\label{eq:cumulative_work}
  \ave{W(t)}_{v_0} = \int_0^t\plaind{t'}\  \dot{\lambda}_{v_0}(t') \left( \lambda_{v_0}(t') - \ave{x(t')}_{v_0} \right) \ . 
\end{equation}
The cumulative work exhibits positive jumps as a result of the initial jumps in the protocols (except for the special case shown as example 3 in the figure, where the initial jump in the protocol vanishes), followed by a transient regime of decreasing cumulative work up to $t \lesssim \tau$. For $t \gtrsim \tau$, the cumulative work either flattens or increases again since, by then, the self-propulsion is decorrelated from the initial value $v_0$ and, thus, all extractable work has been extracted. 
This leads to differing values of the total average work $\ave{W}_{v_0} = \ave{W(\tf)}_{v_0}$ 
\begin{align}
\begin{split}
  \label{eq:avg_work_partialconditional}
  \ave{W}_{v_0} 
  &= \frac{d^2_{v_0}}{\tf + 2/k} 
  -\frac{k v_0^2}{2(k+1/\tau)^2} 
  - {\frac{\tau v_0^2}{8}} \left(1- e^{-2\tf/\tau} \right)  \ .
\end{split}
\end{align}
The total work is positive in examples 1 and 2, zero in example 3, and negative in example 4 (meaning work extraction), as shown in the black squares in the bottom panels of Fig.~\ref{fig:work_incs}(a).
(Note that in example 3, the parameters are fine-tuned to produce both a vanishing initial jump and zero total work, but these phenomena are independent.)
Notably, there is a large parameter region of work extraction, as highlighted by the red region in Fig.~\ref{fig:work_incs}(c) which marks $\ave{W}_{v_0} < 0$.
This region of work extraction appears for sufficiently large magnitudes of $v_0$ as $\langle W \rangle_{v_0}$ is quadratic in $v_0$. 
The maximum in $v_0$ does not occur at $v_0 = 0$ because the target $\lambdaf\neq0$ breaks spatial symmetry with respect to the direction of self-propulsion motion.
This means that positive and negative values of the initial orientation $v_0$ are no longer energetically equivalent. 
As a result, it requires smaller magnitudes of $v_0$ to extract work when the self-propulsion aligns with the target ($v_0 > 0$ and $\lambdaf > 0$, as in example 4) compared to when they are anti-aligned ($v_0 < 0$ but $\lambdaf > 0$, as in example 1). 
It is easier to extract work when the measurement $v_0$ and target $\lambdaf$ align. 

As $\tau$ increases, this alignment effect becomes more pronounced. 
With longer persistence times, the particle maintains its orientation for a longer duration, which can be energetically beneficial or unfavorable depending on the alignment of target and self-propulsion.
When $v_0 > 0$ aligns with $\lambdaf$, a longer persistence time means that the particle continues to actively propel itself in the favorable direction for a more extended period. This sustained orientation reduces the external work required to move the particle to the target, making it energetically easier to extract work from the system.
In contrast, for $v_0 < 0$, when the self-propulsion opposes $\lambdaf$, increasing $\tau$ prolongs the unfavorable alignment, making it more difficult to extract work. 
In the ballistic limit, $\tau \to \infty$, the region of work extraction for negative $v_0$ disappears entirely.
Taking this limit of Eq.~\eqref{eq:avg_work_partialconditional}, the average work
\begin{equation}\label{eq:W_v0_tau_limit}
  \lim_{\tau\to\infty} \ave{W}_{v_0} =  \frac{\lambdaf^2}{\tf + 2/k} - v_0 \lambdaf \ ,
\end{equation}
reduces from a quadratic to a linear dependence in $v_0$. In this limit, work can only be extracted for large enough values of $v_0$, specifically when $v_0 > \lambdaf/(\tf + 2/k)$.

\begin{figure*}
    \centering
    \includegraphics[width=\linewidth]{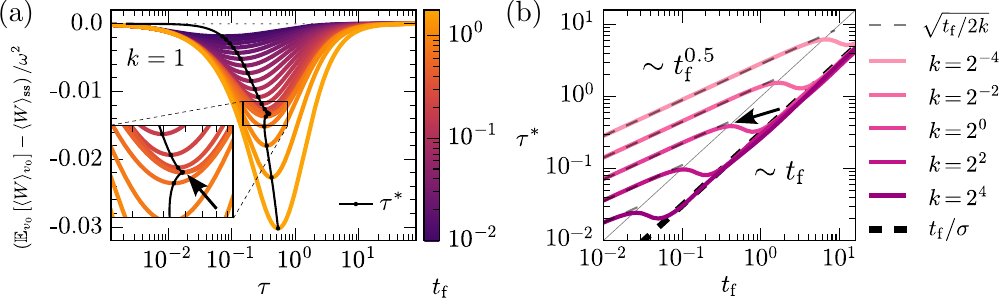}
    \caption{
    (a) The average closed-loop work $\E_{v_0}\! [ \ave{W}_{v_0} ]$, in Eq.~\eqref{eq:avgavg_work_partialconditional}, shows a pronounced minimum for a finite persistence time $\tau$, here shown for $k=1$ and $\tf \in [10^{-2},10^{0.5}]$. The $y$-axis shows the reduction in average work compared to the passive limit ($\tau\to0$). Inset shows the kink (arrow) in the minimum $\tau^*$ (black line) as a function of $\tf$. 
    (b) Optimal persistence time $\tau^*$ as a function of $\tf$ for various $k$ shows the kink (arrow) in $\tau^*$ is the result of a crossover of timescales as $\tf$ increases. Thin dashed line represents the $\tf \ll 1/k$ limit in which $\tau^* = \sqrt{\tf/2k}$, thick dashed line the opposite limit $\tf \ll 1/k$ in which $\tau^* = \tf/\sigma$ with $\sigma\approx 3.2453$. 
    }
    \label{fig:avg-work}
\end{figure*}

The possibility to extract work in $\ave{W}_{v_0}$ is the result of a competition between the three terms in Eq.~\eqref{eq:avg_work_partialconditional}. The first term represents a cost, which is counterbalanced by the two extraction terms, similar to the case of $\ave{W}_{x_0,v_0}$ in Eq.~\eqref{eq:avg_work_conditional}.
In the first extraction term, measuring $v_0$ provides information about the average initial position $\xvo = \ave{x(0)}_{v_0}$. Thus, similarly to the case with $x_0$ measurement discussed for $\ave{W}_{x_0,v_0}$ in Eq.~\eqref{eq:avg_work_conditional}, the $v_0$ measurement allows for the extraction of some of the potential energy stored in the expected initial configuration.
The second extraction term represents conversion of active energy from the self-propulsion to work. After measuring $v_0\not=0$, setting the trap to a position where the particle actively \textit{climbs up} the potential, and then moving the trap along the direction of self-propulsion, it is possible to extract net work from the activity until the orientation decorrelates. 
In contrast to the work extracted from the potential energy, this operation requires a finite process duration, $\tf>0$. Neither mechanism requires an ambient temperature, $D>0$, in contrast with work extraction from potential energy through positional measurements \cite{abreu2011extracting}.

\subsubsection{Work averaged over measurements}

Averaging over the measurement outcomes $v_0$, we find that the average work of the closed-loop protocol in Eq.~\eqref{eq:lambdaoptactive} is
\begin{align}\label{eq:avgavg_work_partialconditional}
\begin{split}
  \E_{v_0}\!\left[ \ave{W}_{v_0} \right] &= 
    - \frac{k \omega^2}{2(k+1/\tau)^2} 
    - \frac{1}{8} \tau \omega^2 \left(1- e^{-2\tf/\tau} \right) 
    \\ 
    &+ \frac{\lambdaf^2 + \frac14 \omega^2 \left(\frac{2}{k+1/\tau} + \tau (1 - e^{-\tf / \tau})\right)^2}{\tf + 2/k} 
    \ .
\end{split}
\end{align}
We decompose the average work into a contribution that also arises for open-loop control, i.e., the open-loop work $\ave{W}_\op$, and a (dimensionless) excess work $\Delta w$
\begin{align}\label{eq:avgavg_work_decomposition}
  \E_{v_0}\!\left[ \ave{W}_{v_0} \right] &= \ave{W}_\op - \tau \omega^2 \Delta w 
    \ .
\end{align}
The excess work 
\begin{align}
\begin{split}\label{eq:norm_excess_work}
  \Delta w := & -\frac1{\tau \omega^2} \left( \E_{v_0}[\ave{W}_{v_0}] - \ave{W}_\op\right)
  \\
  =&
   \frac12 \frac{k\tau}{(1 + k\tau)^2}
  + \frac18  \left(1- e^{-2\tf/\tau} \right) 
  \\
  &- \frac14  \frac{\left( \frac{2}{1 + k\tau} + (1 - e^{-\tf / \tau})\right)^2 }{\tf/\tau + 2/(k\tau)} 
\end{split}
\end{align}
captures the average energetic effects of the closed-loop control compared to the open-loop control. 
Importantly, the excess work is non-negative, $\Delta w \geq 0$, which implies that the average work is bounded from above by the open-loop expression $\ave{W}_\op$,
\begin{align}\label{eq:pce_nonpositive}
  \E_{v_0}\!\left[\ave{W}_{v_0}\right] \leq \ave{W}_\op \ .
\end{align}
This bound shows that, on average, the protocol in Eq.~\eqref{eq:opt_partialconditional} reduces the required work compared to the open-loop case. 
Figure~\ref{fig:Wv0_tau} illustrates the energetic benefit of the additional information about the self-propulsion for different values of persistence time.

Remarkably, $\Delta w$ vanishes for $\tau\to0$ and $\tau\to\infty$, such that $\E_{v_0}\![\ave{W}_{v_0}]$ reduces to the open-loop expression in \emph{both} of these limits,
\begin{equation}\label{eq:tau_limit_avgW}
\begin{split}
  \lim_{\tau\to \infty} \E_{v_0}\!\left[ \ave{W}_{v_0} \right] &= \lim_{\tau\to0} \E_{v_0}\!\left[ \ave{W}_{v_0} \right] 
  \\
  &=  \frac{\lambdaf^2}{\tf + 2/k}  \equiv \ave{W}_\op \ ,
\end{split}
\end{equation}
as shown in Fig.~\ref{fig:Wv0_tau}. 
While this result is expected for $\tau\to0$, when the active particle reduces to a passive one, it may seem counterintuitive at first for $\tau\to\infty$. In this limit, the particle orientation never relaxes, seemingly providing an ``infinite reservoir'' of activity that could be extracted, reducing the work cost. To see why this is not the case, consider the ballistic limit of the average work per measurement in Eq.~\eqref{eq:W_v0_tau_limit}.
In this limit, the average work becomes linear in $v_0$, and the linear term vanishes when taking an average over $v_0$, so that $\E_{v_0} [ \frac{\lambdaf^2}{\tf + 2/k} - v_0 \lambdaf ] =  \frac{\lambdaf^2}{\tf + 2/k} = \ave{W}_\op$.
Physically, an orientation $v_0$ that points in the same direction as $\lambdaf$ reduces the associated work by $-v_0 \lambdaf$, while the opposite case increases the work cost by a term of \textit{equal magnitude} (namely $v_0 \lambdaf$); thus, these contributions cancel out on average.

\subsubsection{Optimal persistence time $\tau$}

An intriguing consequence of the equal limits in Eq.~\eqref{eq:tau_limit_avgW} is the existence of a persistence time $\tau$ that minimizes the average work 
\begin{equation}\label{eq:tau_min}
  0 < \tau^* = \underset{\tau}{\argmin} \ \E_{v_0}\!\left[ \ave{W}_{v_0} \right] < \infty \ .
\end{equation}
The optimal $\tau^*$ is finite for finite $k$ and $\tf$ and results in a pronounced minimum in $\E_{v_0}\![\ave{W}_{v_0}]$, as visible in Fig.~\ref{fig:Wv0_tau}. 
Figure~\ref{fig:avg-work}(a) illustrates the optimal persistence time $\tau^* = \tau^*(\tf,k)$ as a function of $\tf$ for $k=1$. There is a noticeable kink in $\tau^*$ at intermediate values of $\tf$, marked by the arrow in the inset. 
This kink results from a crossover between two distinct scalings in the system, as detailed in Fig.~\ref{fig:avg-work}(b), which shows $\tau^*$ as a function of $\tf$ for different $k$ values. 
Upon closer inspection, $\tau^*$ scales with the square-root of $\tf$ for protocol durations shorter than the relaxation period of the trap, $\tf \ll 1/k$.
Specifically, expanding $\E_{v_0}\!\left[ \ave{W}_{v_0} \right]$ in Eq.~\eqref{eq:avgavg_work_partialconditional} for small $\tf$ 
\begin{equation}
  \E_{v_0}\!\left[\ave{W}_{v_0}\right] = \ave{W}_\op - \frac{\tf}{4(1+k\tau)^2}  + \frac{(2+3k\tau)\tf^2}{8\tau(1+k\tau)^2}  + O(\tf^2)
\end{equation}
and solving for the optimal $\tau$ yields 
\begin{equation}
  \tau^* = \sqrt{\frac{\tf}{2k}} \left(1 + O\left(\sqrt{k\tf}\right)\right) \ .
\end{equation}
In this regime, the protocol duration is shorter than the average time for the particle to relax in the trap, limiting the protocol's ability to extract active energy. To counter this effect, the optimal persistence time $\tau^*$ increases relative to $\tf$ which enables the protocol to exploit more of the activity, despite the inability to fully relax in the harmonic trap. 
As $\tf$ exceeds the relaxation time, $\tf \gg 1/k$, there is a transition to a linear dependence of $\tau^*$ on $\tf$.
To make this precise, we first take the limit $k\to\infty$ in $\E_{v_0}\! [ \ave{W}_{v_0} ]$ which effectively removes the square-root regime in $\tau^*$ by pushing the crossover towards $\tf \sim 1/k \to 0$. We then optimize with respect to $\tau$, which leads to the transcendental equation $0=\Phi(\tf/\tau)$, where
\begin{equation}\label{eq:phi}
  \Phi(x) = 4-x-4(2+x)e^{-x}+(2x^2+5x+4)e^{-2x} \ .
\end{equation}
The large $\tf$ limit is thus $\tau^* = \tf/\sigma$, where the inverse slope $\sigma$ solves $0 = \Phi(\sigma)$ and is approximately $\sigma \approx 3.2453$. 
In this regime, the protocol duration $\tf$ is sufficient to allow full relaxation within a single execution. 
The optimal persistence time $\tau^*$ now represents the following balance: a longer persistence time enables the system to extract work from favorable directions of $v_0$, but excessively long persistence times lead to domination by unfavorable directions of $v_0$, as discussed around Eq.~\eqref{eq:W_v0_tau_limit}.
Notably, in this regime of $\tf \gg 1/k$, the ratio $\tau / \tf = 1/\sigma$ is constant. Hence, when the protocol duration allows full relaxation within a single execution, the optimal persistence time $\tau^*$ is approximately \emph{one third} of the protocol duration $\tf$.
The aforementioned shift from a square-root to a linear scaling is responsible for the kink observed in Fig.~\ref{fig:avg-work}(a).

The existence of a finite optimal $\tau^*$ means that the cost to translate an active self-propelled particle is lower than for a passive particle ($\tau \to 0$) or a ballistic particle ($\tau \to \infty$). As a result, a finite persistence time, representing non-equilibrium fluctuations, offers an advantage in the closed-loop protocol.
This result highlights that the closed-loop control protocol is most effective for physically realistic active models with finite persistence time.


\subsection{The impact of measurement uncertainty}
\label{sec:epsilon}

Every measuring device is inherently subject to a certain degree of noise, error, and uncertainty. In this section, we quantify the impact of such measurement uncertainties on the performance of the control. Specifically, we calculate the work correction associated with the uncertainty. To do so, we first compute the joint and conditional probability densities of the true and measured system states.

We introduce the quantities $\vgt$ and $\vm$ to denote the true and measured values of $v(0)$, respectively. As before, we focus on the case with a measurement of $v(0)$ only. 
We introduce a new ensemble consisting of the collection of all realizations of trajectories with initial condition $v(0) = \vgt$ and with the additional stochasticity arising from the uncertainty in the measurement $\vm$ which is the value used by the controller to determine the protocol $\lambda_{v_0}$ in Eq.~\eqref{eq:lambdaoptactive}. 
We define the corresponding notation for the average in this ensemble as
\begin{equation}
  \ave{\bullet}_{\vgt,\vm} := \ave{\bullet | v(0) = \vgt, \text{measurement taken is } \vm } \ .
\end{equation}

We assume that the measurement has a Gaussian-distributed error $\epsilon$, such that the (post-measurement) probability density function $P(\vm|\vgt)$ of the measurement outcome $\vm$ conditioned on the true value $\vgt$ is a normal distribution with variance $\epsilon^2$ centered at the true value $\vgt$.
In App.~\ref{app:PDFs}, we discuss the new ensemble $\ave{\bullet}_{\vgt,\vm}$ in further detail and calculate the joint distribution $P(\vm, \vgt)$ for AOUPs and RTPs.

We now consider the case where the controller applies the protocol $\lambda_{v_0}$ in Eq.~\eqref{eq:lambdaoptactive} based on the measurement outcome $\vm$ instead of the real value $\vgt$. 
If the controller has additional knowledge about the measurement uncertainty, it is possible to further reduce the energetic cost by adjusting the protocol to the level of uncertainty, as discussed in Refs. \cite{abreu2011extracting,cocconi2023optimal}, but we do not consider this here.

The protocol that the controller applies is
\begin{align}\label{eq:lambdam}
\begin{split}
    \lambdam(t) &:= \frac{\vm}{k+1/\tau}
  + \frac{d^{\rm m}_{\vm}}{\tf + 2/k} t  + \frac{\tau \vm}{2}  (1-e^{-t/\tau})
  \\
  &+ \frac{d_{\vm}^{\rm m}}{k\tf + 2} - \frac{\vm}{2 k} e^{-t/\tau}
  \ ,
\end{split}
\end{align}
with $\lambdam(0) = 0, \lambdam(\tf) = \lambdaf$ and
where $d_{\vm}^{\rm m}$ are obtained from $d_{v_0}$ by substituting $v_0$ by $\vm$
\begin{align}\label{eq:d_vm^m}
  d_{\vm}^{\rm m} &= \lambdaf - \frac{\vm}{k+1/\tau} - \frac{\tau \vm}{2}  \left(1-e^{-\tf/\tau}\right) \ .
\end{align}


The protocol $\lambdam$ leads to an additional energetic cost compared to the optimal protocol in Eq.~\eqref{eq:opt_partialconditional} due to the measurement uncertainty, which we quantify in the following. 
The total true average work accounting for the measurement uncertainty can be obtained by evaluating the work functional in Eq.~\eqref{eq:workfunctional} using $\lambdam$ and the average particle position under $\lambdam$ which we denote by $\ave{x(t)}_{\vgt,\vm}^{\rm r}$
\begin{equation}
\begin{split}
\label{eq:Wtotal_SM}
  \ave{ W }_{\vgt,\vm}^{\rm r} &:= W[\lambdam(t), \ave{x(t)}_{\vgt,\vm}^{\rm r}] 
  \\
  &= k \int_0^{\tf}\plaind{t}\ \dot{\lambda}^{\rm m} \left(\lambdam - \ave{x}_{\vgt,\vm}^{\rm r} \right) \ .
\end{split}
\end{equation}
In App.~\ref{app:Work-measurement-uncertainty}, we derive the average total work as
%
\begin{align}\label{eq:avgavg_work_w_uncertainty}
  \E_{\vgt,\vm}[ \ave{W}_{\vgt,\vm}^{\rm r} ] &= \ave{W}_\op - (\omega^2 - \epsilon^2) \tau \Delta w \ ,
\end{align}
where the excess work is defined in Eq.~\eqref{eq:norm_excess_work}. This result should be compared with Eq.~\eqref{eq:avgavg_work_decomposition}.
%
%
Equation~\eqref{eq:avgavg_work_w_uncertainty} states that the average work in the presence of uncertainty is \emph{increased} relative to the ideal closed-loop case, by a term proportional to $+\epsilon^2$. This additional cost diminishes the work reduction achieved through closed-loop control (the factor proportional to $-\omega^2$). 
However, as long as the measurement uncertainty $\epsilon$ remains smaller than the self-propulsion variability $\omega$, the net benefit of closed-loop control persists, in the sense that the average work remains lower than in the open-loop case; increasing $\epsilon$ gradually reduces this advantage.


\subsection{Cost of information acquisition}\label{app:info}

\begin{figure}[tb]
    \centering
    \includegraphics[width=0.8\columnwidth]{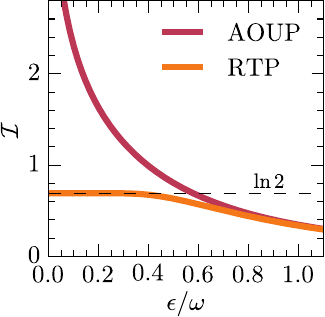}
    \caption{Mutual information, Eq.~\eqref{eq:MinfoAOUP}, as a function of measurement uncertainty $\epsilon$. 
    Dashed lines show analytical estimates for mutual information, Eq.~\eqref{eq:Mutual-info-RTP-result} for an RTP assuming $\epsilon/ \omega \ll 1$.
    }
    \label{fig:mi}
\end{figure}

For systems subject to closed-loop control, the framework of information-thermodynamics has established that
the fundamental limit to the \textit{thermodynamic cost of the measurement} is given by the reduction of uncertainty about the system state through that measurement, scaled with the temperature~\cite{parrondo2023thermodynamics,parrondo2015thermodynamics,sagawa2010generalized,sagawa2008second,cao2009thermodynamics,abreu2011extracting,paneru2022colossal,horowitz2010nonequilibrium}. 
The reduction of uncertainty about the actual system, $z$, state by a measurement with outcome, $z_\mathrm{m}$, is quantified by the mutual information
\begin{align}
\elabel{Minfo_def}
  \! \mathcal{I}[z_\mathrm{m}; z] &= H(z)-H(z_\mathrm{m}|z)
=  \E_{z,z_\mathrm{m}} \!\!\left[ \ln\frac{P(z,z_\mathrm{m})}{P(z)P(z_\mathrm{m})} \!\right] \!,
\end{align}
where $H$ denotes the Shannon entropy. This holds under the assumption that both the measurement device or ``demon" and controlled system operate at the same temperature. To embed our setup into this framework, we assume here and in the following that the demon operates at temperature $T=D_v$.

In our closed-loop control scheme, the demon measures $v_0$. Due to the correlations present in the non-equilibrium steady state, the measurement of $v_0$ however also reduces the uncertainty about $x_0$, so that the measurement cost is given by
  $ D_v\mathcal{I}[\vm; \vgt,x_0]$, where we denote by $\vm$ the measurement outcome and by $\vgt$ the actual system state. 
  Using the chain rule for mutual information, we find that
\begin{equation}
    \mathcal{I}[x_0 ,\vgt; \vm] = \mathcal{I}[x_0 ; \vm | \vgt] + \mathcal{I}[\vgt; \vm]\,.
\end{equation}
The measurement $\vm$ is independent of $x_0$, see Eq.~\eqref{eq:measurementassump}, which implies that $p(\vm|x_0,\vgt)=p(\vm|\vgt)$ and as a result $\mathcal{I}[x_0; \vm | \vgt] = 0$.
It follows that the lower bound to the information acquisition cost is $D_v \mathcal{I}[\vm; \vgt]$.

To evaluate $  \mathcal{I}[\vm; \vgt]$ for AOUPs and RTPs, 
we insert into Eq.~\eref{Minfo_def} the joint and marginalized probability densities [stated in App.~\ref{app:PDFs}, see Eqs.~\eqref{eq:joint-AOUP},~\eqref{eq:marginal-AOUP},~\eqref{eq:marginal-RTP}, and \eqref{eq:joint-RTPs}].
Using these results, we find for AOUPs,
\begin{align}
\label{eq:MinfoAOUP}
&  \mathcal{I}_\mathrm{AOUP}[\vm;\vgt] =
 \iint_{\mathbb{R}} \plaind{\vgt}\plaind{\vm}\, P(\vgt,\vm)\ln\frac{P(\vgt,\vm)}{P(\vgt)P(\vm)} 
\nonumber \\
 &=
 \iint_{\mathbb{R}} \plaind{\vgt}\plaind{\vm}
 \mathcal{N}_{\vgt}(0,{\omega}^2)  
 \mathcal{N}_{\vm}(\vgt,\epsilon^2)  
 \ln\frac{\mathcal{N}_{\vm}(\vgt,\epsilon^2)}{ \mathcal{N}_{\vm}(0,{\omega}^2+\epsilon^2) }
\nonumber \\
 &=
 \frac{1}{2}\ln\left(1+\frac{{\omega}^2}{\epsilon^2}\right)\,.
\end{align}
This result is in qualitative agreement with the corresponding results for positional measurements on PPs~\cite{abreu2011extracting}, which are also described by Gaussian densities.
Figure~\ref{fig:mi} illustrates the mutual information from Eq.~\eqref{eq:MinfoAOUP} as a function of $\epsilon/\omega$.
As expected, the mutual information vanishes as $\epsilon \to \infty$, where measurements become uninformative, and diverges for $\epsilon \to 0$. This divergence is expected due to the infinite information content in a continuous variable, $v_0 \in \mathbb{R}$ \cite{PolyanskiyWu2014,Gray2011}. 

For RTPs, we make use of Bayes' theorem to evaluate the mutual information \eref{Minfo_def}, from which we obtain
\begin{widetext}
\begin{align}
 \mathcal{I}_\mathrm{RTP}[\vm ;\vgt]&=
-\iint_{\mathbb{R}} P(\vgt,\vm) \ln\frac{P(\vgt)}{P(\vgt|\vm)} \plaind{\vgt}\plaind{\vm}
 \nonumber  \\
    &
    =
  \iint_{\mathbb{R}}  \mathcal{N}_{\vm}(\vgt,\epsilon^2) \frac{1}{2} [\delta(\vgt+{\omega})+\delta(\vgt-{\omega})]
 \ln\frac{2\mathcal{N}_{\vm}(\vgt,\epsilon^2)}{\mathcal{N}_{\vm}(-{\omega},\epsilon^2)+
\mathcal{N}_{\vm}({\omega},\epsilon^2)}
\plaind{\vgt}\plaind{\vm} 
 \nonumber  \\
    &=-\frac{1}{2}
  \int_{\mathbb{R}}  \mathcal{N}_{\vm}({\omega},\epsilon^2) 
 \ln\frac{\mathcal{N}_{\vm}(-{\omega},\epsilon^2)+
\mathcal{N}_{\vm}({\omega},\epsilon^2)}{2\mathcal{N}_{\vm}({\omega},\epsilon^2)}\mathrm{d}\vm 
  -\frac{1}{2} \int_{\mathbb{R}}  \mathcal{N}_{\vm}(-{\omega},\epsilon^2) 
 \ln\frac{\mathcal{N}_{\vm}(-{\omega},\epsilon^2)+
\mathcal{N}_{\vm}({\omega},\epsilon^2)}{2\mathcal{N}_{\vm}(-{\omega},\epsilon^2)}\mathrm{d}\vm
 \nonumber  \\
    &
    =-
  \int_{\mathbb{R}}  \mathcal{N}_{\vm}({\omega},\epsilon^2) 
 \ln\left[\mathcal{N}_{\vm}(-{\omega},\epsilon^2)+
\mathcal{N}_{\vm}({\omega},\epsilon^2)\right]\mathrm{d}\vm 
+
  \int_{\mathbb{R}}  \mathcal{N}_{\vm}({\omega},\epsilon^2) 
 \ln\left[{2\mathcal{N}_{\vm}({\omega},\epsilon^2)}\right]\mathrm{d}\vm
\nonumber  \\
    &
    =-
  \int_{\mathbb{R}}  \mathcal{N}_{\vm}({\omega},\epsilon^2) 
 \ln\left[\mathcal{N}_{\vm}(-{\omega},\epsilon^2)+
\mathcal{N}_{\vm}({\omega},\epsilon^2)\right]\mathrm{d}\vm 
  -
  \frac{1}{2}\left[ 1+ \ln(\pi \epsilon^2/2)
 \right] 
 \ .
 \label{eq:Mutual-info-RTP-intermediate-step}
\end{align}
\end{widetext}
The final integral in \Eref{Mutual-info-RTP-intermediate-step} cannot be evaluated in closed form in general. However, assuming $\epsilon$ is small relative to ${\omega}$, i.e., $\epsilon\ll|\omega|$, the term $\mathcal{N}_{\vm}({\omega}, \epsilon^2)$ becomes negligible near $\vm = 0$. Consequently, the probability of measuring $P(|\vm| = 0)$ vanishes, which is reasonable for an RTP. Under this assumption, the logarithm in Eq.~\eqref{eq:Mutual-info-RTP-intermediate-step} can be approximated by $\ln (2)$, allowing the integral to be evaluated as
\begin{align}
& \int_{\mathbb{R}}  \mathrm{d}\vm\, \mathcal{N}_{\vm}({\omega},\epsilon^2) 
 \ln\left[\mathcal{N}_{\vm}(-{\omega},\epsilon^2)+
\mathcal{N}_{\vm}({\omega},\epsilon^2)\right]
\nonumber
\\
&\approx
\int_{\mathbb{R}}  \mathrm{d}\vm\, \mathcal{N}_{\vm}({\omega},\epsilon^2) 
 \ln\left[\mathcal{N}_{\vm}({\omega},\epsilon^2)\right]
\nonumber 
\\
& = -\frac{1}{2}\left[ 1+ \ln(2\pi \epsilon^2)
 \right]\,.
 \end{align}
Together we find the information cost for RTPs for vanishing error
\begin{equation}\label{eq:Mutual-info-RTP-result}
     \mathcal{I}_\mathrm{RTP}[\vm;\vgt]
    \simeq \ln(2) \ .
\end{equation}
As might be expected, the measurement removes \textit{one bit} of uncertainty for RTPs, regardless of a (small) measurement uncertainty. This expression becomes exact for $\epsilon \to 0$.

Figure~\ref{fig:mi} shows $ \mathcal{I}_\mathrm{RTP}[\vm; \vgt]$ obtained via numerical integration of the exact expression in Eq.~\eqref{eq:Mutual-info-RTP-intermediate-step}. As evident from the figure, the estimate $\ln(2)$ remains accurate in a reasonable range of small $\epsilon$. For large $\epsilon$,
the difference in the distributions of $v_0^r$ for RTP and AOUP become negligible. In this regime, $\mathcal{I}_\mathrm{RTP}[\vm; \vgt]$ can be approximated by $\mathcal{I}_\mathrm{AOUP}[\vm; \vgt]$ from Eq. \eqref{eq:MinfoAOUP}.

Comparing Eqs.~\eqref{eq:Mutual-info-RTP-result} and \eref{MinfoAOUP} shows that the measurement cost is generally lower for RTPs than for AOUPs, as expected, while it approaches zero as $\epsilon \to \infty$. Numerical integration of Eq. \eqref{eq:Mutual-info-RTP-intermediate-step} suggests that this holds true for all values of $\epsilon$,
\begin{align}
    \mathcal{I}_\mathrm{RTP}[\vm;\vgt]\leq \mathcal{I}_\mathrm{AOUP}[\vm;\vgt] \ ,
\end{align}
as illustrated in Fig.~\ref{fig:mi}.

\section{Information engine}\label{sec:infoengine}

A notable application of optimal closed-loop protocols are periodic machines. Based on our results, it is possible to construct a minimal, optimal information engine that harvests energy from the activity using self-propulsion measurements. 
In the subsections \ref{sec:engineprotocol} and \ref{sec:enginecycle}, we revert back to the case of error-free measurements, while the measurement uncertainty will be dealt with in the subsequent subsection.

\subsection{Optimal protocol}
\label{sec:engineprotocol}

\begin{figure*}
    \includegraphics[width=\textwidth]{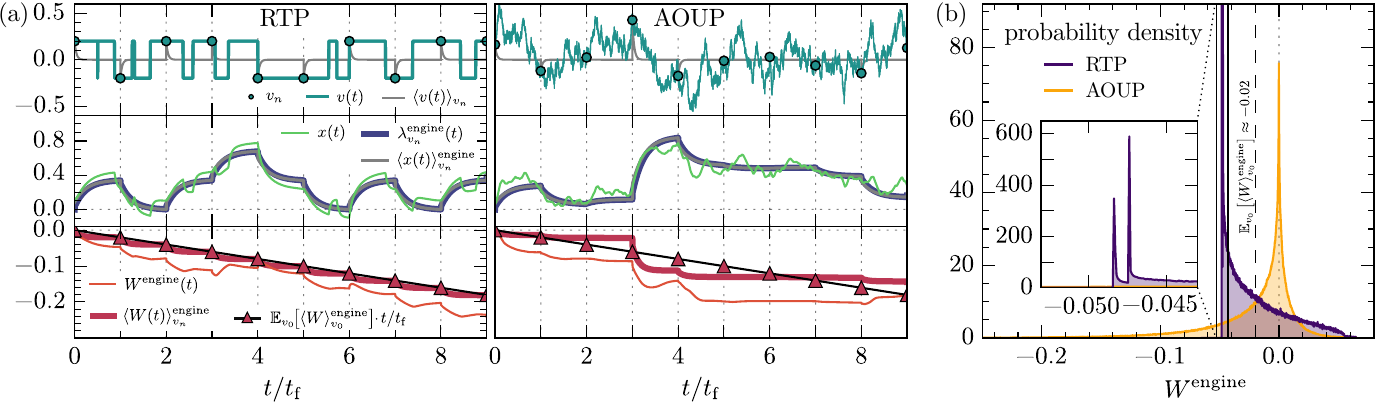}
    \caption{
    Active information engine: 
    (a) Typical trajectories of the information engine for an RTP (left) and an AOUP (right). From top to bottom: trajectories of the self-propulsion, particle position, trap position and accumulated work 
    $\ave{W(t)}_{v_0}^{\rm engine} = k \int_0^t\plaind{t'}\  \dot{\lambda}^\text{engine}_{v_0}(\lambda^\text{engine}_{v_0} - x)$. Self-propulsion measurements $v(n\tf)$ are indicated with circles at the beginning of each cycle, and they are assumed to have no error ($\epsilon = 0$). 
    (b) Work distribution per cycle for AOUP and RTP, Eq.~\eqref{eq:work_engine_percycle_steady}, from simulations. The mean is identical for AOUP and RTP and matches the theory (dashed line).
    Parameters in all panels: $D=0$, $k = 2$, $\tf = 10$, $\omega = 0.2$, $\tau = \tau^* \approx 2.615$.
    }
    \label{fig:engine}
\end{figure*}

Starting from $\ave{W}_{v_0}$ in Eq.~\eqref{eq:avg_work_partialconditional}, we perform a secondary optimization with respect to $\lambdaf$, from which we obtain the optimal shifting distance after measuring $v_0$ 
\begin{equation}\label{eq:lambdafengine}
  \lambdaf^\engine = \xvo + \frac{\tau v_0}2  (1 - e^{-\tf / \tau}) \ ,
\end{equation}
which sets $d_{v_0}=0$, thereby eliminating the cost term in $\ave{W}_{v_0}$. 
Substituting this optimal distance into Eq.~\eqref{eq:opt_partialconditional} yields the (information) engine protocol and associated mean particle position
\begin{subequations}\label{eq:opt_engine}
\begin{align}
  \ave{x(t)}^\engine_{v_0} &= \xvo  + \frac{\tau v_0}{2}  (1-e^{-t/\tau})  \ ,
  \\
  \lambda^\engine_{v_0}(t) &=
  \begin{cases}
      0 & t=0 \\
      \ave{x(t)}_{v_0}^\engine - \frac{v_0}{2 k} e^{-t/\tau} & 0 < t < \tf \\
      \xvo + \frac{\tau v_0}2  (1 - e^{-\tf / \tau}) & t =\tf
  \end{cases}
  \ .
\end{align}
\end{subequations}
At $t = \tf$, the protocol resets to the mean particle position, $\lambda_{v_0}^\engine(\tf) = \ave{x(\tf)}_{v_0}^\engine$, unlike previous protocols 
where $\lambda_{v_0}(\tf) = \lambdaf \neq \ave{x(\tf)}_{v_0}$. 
This allows the protocol to extract any potential energy remaining at $t=\tf$, which the optimal protocol for closed-loop control with boundary condition $\lambda_{v_0}(\tf) = \lambdaf$ cannot harness, due to the boundary constraint.

We note that the protocol $ \lambda^\engine_{v_0} $ shares conceptual similarities with the protocols considered in Refs.~\cite{cocconi2024efficiency, malgaretti2022szilard}. In those works, the protocols involve an external force applied directly to the particle, whereas in our case, the external force acts on the particle indirectly through the harmonic trap.
The resulting force on the particle in our system is half of the average self-propulsion,
\begin{equation}
  F(t) = k\left(\lambda^\engine_{v_0}(t) - \ave{x(t)}_{v_0}^\engine\right) = -\frac{\ave{v(t)}_{v_0}}{2}  \ .
\end{equation}  
Remarkably, despite the different setup, this strategy qualitatively resembles the one in Ref.~\cite{cocconi2024efficiency}, where the instantaneous self-propulsion replaces the average value due to the use of continuous measurements. This continuous measurement protocol can be obtained from ours in the limit of vanishing protocol duration $\tf\to0$.

The average work of the engine protocol per measurement is
\begin{equation}
\begin{split}
  \ave{W}_{v_0}^\text{engine} 
  &=  
    - \frac12 \frac{k v_0^2}{(k + 1/\tau)^2}
    - \frac{\tau v_0^2}8   \left(1- e^{-2\tf/\tau} \right) \ ,
\end{split}
\end{equation}
which confirms that the engine protocol indeed extracts work for any measurement outcome $v_0 \not = 0$
\begin{equation}
  \ave{W}_{v_0}^\text{engine} \leq 0 \ ,
\end{equation}
and for all parameter values $k,\tau,\tf$. 
As a result, the average work
\begin{equation}\label{eq:avg_work_engine}
\begin{split}
  \E_{v_0}\!\left[ \ave{W}_{v_0}^\text{engine} \right] &=  
    - \frac12 \frac{k \omega^2}{(k + 1/\tau)^2}
    - \frac{\tau \omega^2}8  \left(1- e^{-2\tf/\tau} \right)
    \ ,
\end{split}
\end{equation}
is also non-positive. It reaches its minimum $-\tau \omega^2/4$ for $k\tau = 1$ in the quasistatic limit.

\subsection{Repeated execution of engine protocol}
\label{sec:enginecycle}

By iterating the protocol $\lambda^\engine_{v_0}$, we obtain an engine with period $\tf$.
To execute the $n$-th cycle, we apply the following procedure:
we measure the self-propulsion, $v_n = v(n \tf)$ at the start of the cycle, and then apply the protocol $\lambda^\engine_{v_n}(t)$.

The engine builds up correlations over time. 
Specifically, the sequence of measurements $v_0, v_1, v_2, \ldots$ forms a Markov chain, where each measurement $v_n$ for $n > 0$ is drawn from a state that depends on the previous measurement.
Because $v(t)$ does not fully relax between measurements, 
the expected particle position given a measurement for $n>0$  now becomes
\begin{align}
\begin{split}
  x_{v_n} &:= 
    \frac{ \Cov_{x_0,v_0} x(\tf) v(\tf) }{\Var_{x_0,v_0}(\tf)} (v_n - \E_{v_n}[v_n | v_{n-1}]) 
   \ ,
\end{split}
\end{align}
which is further discussed in App.~\ref{app:engine_mc}.
In general, $x_{v_n}$ replaces $\xvo$ for $n>0$ in the engine protocol in Eq.~\eqref{eq:opt_engine}.
However, to simplify the following discussion, we assume that the period $\tf$ is much larger than the persistence time $\tau$, making the measurements $v_0,v_1,v_2,\ldots$ effectively uncorrelated. 
In this limit, each measurement $v_n$ is approximately drawn from the stationary state, just like the initial measurement $v_{0}$, and we can continue using the protocol Eq.~\eqref{eq:opt_engine} for all cycles $n>0$.

An example of a repeated execution of the engine protocol in the limit $\tf/\tau \gg 1$ is shown Fig.~\ref{fig:engine}(a) for 9 iterations. The figure also shows simulations of typical stochastic trajectories of the self-propulsion $v(t)$, particle position $x(t)$ and cumulative work $W^\engine(t) = k\int_0^{t}\plaind{t}\, \dot{\lambda}^\engine_{v_n}(t) (\lambda^\engine_{v_n}(t) - x(t))$, along with their average quantities $\ave{v(t)}_{v_n}$, $\ave{x(t)}_{v_n}^\engine$ and $\ave{W(t)}_{v_n}^\engine$, separately for RTPs (left) and AOUPs (right). 

Each iteration extracts an average work of $\E_{v_0}\! [ \ave{W}_{v_0}^\text{engine} ]$.
The distribution of the work per cycle
\begin{equation}\label{eq:work_engine_percycle_steady}
\begin{split}
  W^\engine &= W\left[\lambda_{v_n}^\engine(t), x(t) \right] 
  \\
  &= k\int_0^{\tf}\plaind{t}\, \dot{\lambda}^\engine_{v_n}(t) (\lambda^\engine_{v_n}(t) - x(t))
\end{split}
\end{equation}
differ significantly between RTPs and AOUPs, as shown in Fig.~\ref{fig:engine}(b) using the same parameters as in Fig.~\ref{fig:engine}(a).
Notably, the mode of the distribution is close to 0 and above the mean for AOUPs, but below the mean for RTPs. The work distribution for RTPs is positively skewed, whereas it is negatively skewed for AOUPs. 


{\color{black}
The engine resulting from the periodic execution of the protocol is not strictly cyclic, as over many cycles it exhibits diffusive behavior in both the average particle position and trap center.

Nevertheless, the lack of cyclicity is not a fundamental issue, as it can be resolved by introducing a recovery step with corrections that vanish asymptotically.
Specifically, after every $N$ cycles, which lead to a diffusive displacement of order $\Delta x \sim N^{1/2}$ and work extraction $W\sim N$, the trap center is reset to the origin using a passive optimal protocol [Eq.~\eqref{eq:opt_stationary}] of duration $t_\mathrm{rec}$. 
This results in
a process that is cyclic at the level of average quantities, with period length $N t_\mathrm{f} + t_\mathrm{rec}$.
If the duration is chosen to scale with $t_\mathrm{rec}\sim N^{\alpha}$ with $0<\alpha<1$, the additional work cost in Eq.~\eqref{eq:opt_work_ext_openloop} also scales sublinearly with $W_\mathrm{rec}\sim \frac{N}{N^\alpha +1}\sim N^{1 -\alpha}$. Consequently, corrections to both power output and information efficiency scale sublinearly with $N$ as well (as $\sim N^{1-\alpha}$), and thus remain asymptotically negligible. 
This implies that the results obtained for the engine without a recovery step remain asymptotically equivalent to those of a cyclic engine. 

An alternative strategy is to modify the protocol such that the engine is reset to its initial position at the end of each cycle. 
This is achieved by setting $\lambda_\mathrm{f} = 0$, instead of using the optimal final value in Eq.~\eqref{eq:lambdafengine}.
The resulting protocol is described by Eq.~\eqref{eq:opt_partialconditional} with $\lambdaf=0$. 
While this comes at the cost of reduced work output, a positive net work can still be extracted for sufficiently long cycle durations $t_\mathrm{f}$, as can be seen by substituting $\lambda_\mathrm{f} = 0$ into Eq.~\eqref{eq:avg_work_partialconditional}.
%

Finally, from a pragmatic standpoint, any change to the protocol could in principle be avoided by operating the machine in a periodic spatial range, which bounds delocalization.
}


\subsection{Extractable work with measurement uncertainty}

In this and the following subsections, we address the effect of measurement uncertainty on work extracted by the cyclic engine (using the results from Sec.~\ref{app:Work-measurement-uncertainty}), and the thermodynamic cost of the information  acquisition~\cite{sagawa2010generalized,kullback2013topics,cao2009thermodynamics,parrondo2015thermodynamics}. This allows us to define an engine efficiency as the ratio of average extractable work and the thermodynamic cost.

Substituting $v_0$ by the measurement value $\vm$ in Eq.~\eqref{eq:opt_engine}, we obtain the following protocol
\begin{align}\label{eq:opt_engine_m}\nonumber
  &\lambdameng(t) 
  \\
  &=
  \begin{cases}
      0 & t=0 \\
      \frac{\vm}{k+1/\tau}  + \frac{\tau \vm}{2}  (1-e^{-t/\tau}) - \frac{\vm}{2 k} e^{-t/\tau} & 0 < t < \tf \\
      \frac{\vm}{k + 1/\tau} + \frac{\tau \vm}2  (1 - e^{-\tf / \tau}) & t =\tf
  \end{cases}
  \ .
\end{align}

In a similar spirit to the argument in Sec.~\ref{sec:epsilon}, the total work accounting for the measurement uncertainty under the protocol $\lambdameng$ is 
\begin{align}\label{eq:ndsakldnsakd}
  \ave{ W }^{\engine,\rm r}_{\vgt,\vm} &:= W[\lambdameng(t), \ave{x(t)}_{\vgt,\vm}^{\engine,\rm r}] 
  \nonumber
  \\
  &= k \int_0^{\tf}\plaind{t}\ \dot{\lambda}^{\engine, \rm m}\left(\lambdameng - \ave{x}^{\engine,\rm r}_{\vgt,\vm} \right) \ ,
\end{align}
where $\ave{x(t)}_{\vgt,\vm}^{\engine, \rm r}$ denotes the average particle position under the protocol $\lambdameng$. 

Averaging Eq.~\eqref{eq:ndsakldnsakd}, we find the average work of the engine per cycle
\begin{align}\label{eq:extractable_work_engine}
  \E_{\vgt,\vm}\left[ \ave{ W }^{\engine,\rm r}_{\vgt,\vm} \right] 
  = -(\omega^2-\epsilon^2) \tau \Delta w^{\engine}
  \ ,
\end{align}
with $\Delta w^\engine$ the average work extracted by the engine at $\epsilon\to 0$, non-dimensionalized by division with $\tau \omega^2$,
\begin{equation}\label{eq:deltawengine}
\begin{split}
  \Delta w^\engine &= - \frac{\E_{v_0}\!\left[ \ave{W}_{v_0}^\text{engine} \right]}{\tau \omega^2} 
  \\
  &= \frac12 \frac{k\tau}{(1+k\tau)^2} + \frac{1}{8}\left(1-e^{-2\tf/\tau} \right)  \geq 0 \ . 
\end{split}
\end{equation}
We discuss $\Delta w^\engine$ further in Sec.~\ref{sec:efficiency}.

Thus, the work extraction is reduced by a term proportional to $\epsilon^2$, see Eq.~\eqref{eq:extractable_work_engine}. Moreover, once the measurement error $\epsilon$ exceeds the  magnitude of the standard deviation $|\omega|$, the engine is unable to extract work. For $\epsilon<|\omega|$, work extraction is possible for all finite $k, \tau, \tf$.

\subsection{Information efficiency}
\label{sec:efficiency}

\begin{figure}[hbt]
\centering
\includegraphics[width=0.9\columnwidth]{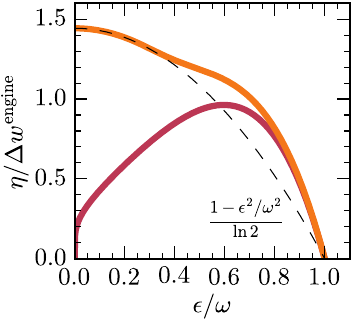}
\caption{Information efficiency $\eta$, \Eref{def_efficiency}, as a function of measurement uncertainty $\epsilon$. The efficiency is shown divided by $\Delta w^\engine$, which is identical for RTPs and AOUPs, in order to obtain a measure that does not depend on the system parameters.
Dashed lines show analytical estimates for the efficiency, Eq.~\eqref{eq:efficiency_detail_RTP}, for an RTP assuming $\epsilon/\omega \ll 1$.
The information efficiency for an AOUP has a maximum at finite measurement error $\epsilon$, whereas it is monotonic decreasing for an RTP.
}
\label{fig:efficiency}
\end{figure}

To obtain an efficiency of the engine, we compare the extractable work in Eq.~\eqref{eq:extractable_work_engine} with the thermodynamic cost of the information acquisition $D_v \mathcal{I}[\vm;\vgt]$, discussed in Sec.~\ref{app:info}. Note that, as before, we only discuss the efficiency of engines with long enough periods to allow for relaxation between cycles. 

The information-efficiency is defined by the mean extracted work per cycle divided by the cost of information acquisition $D_v  \mathcal{I}[\vm;\vgt]$,
\begin{align}
\label{eq:def_efficiency}
\begin{split}
    \eta :=& \frac{-\E_{\vgt,\vm}\left[\ave{W_{\rm m}}_{\vgt,\vm}^\engine \right]}{D_v\,\mathcal{I}[\vm;\vgt]} 
    \\
    =& \frac{1-\epsilon^2/\omega^2}{\mathcal{I}[\vm;\vgt]} \Delta w^\engine \ ,
\end{split}
\end{align}
where we used Eq.~\eqref{eq:extractable_work_engine} in the second equality. Recall that $\Delta w^\engine(k,\tau,\tf)$, given in Eq.~\eqref{eq:deltawengine}, denotes the additional work due to the measurement uncertainty, which is identical for AOUPs and RTPs.

Using \eqref{eq:MinfoAOUP} and \eqref{eq:Mutual-info-RTP-result}, we thus obtain for AOUPs
\begin{subequations}\label{eq:efficiency_detail}
\begin{equation}
     {\eta_{\rm AOUP}} = \frac{2(1-\epsilon^2/\omega^2 )}{ \ln\left(1 + \omega^2/\epsilon^2\right)} {\Delta w^\engine}\ ,
\end{equation}
while for RTPs, in the small $\epsilon$ limit, it approaches
\begin{equation}\label{eq:efficiency_detail_RTP}
\eta_{\rm RTP}\to \frac{1-\epsilon^2/\omega^2 }{ \ln 2}{\Delta w^\engine}   \ . 
\end{equation}
\end{subequations}
Figure~\ref{fig:efficiency} compares $\eta/\Delta w^\engine$ for both RTPs and AOUPs.  
It shows that there is a maximum information-efficiency for finite $\epsilon/\omega$ for AOUPs, meaning that at constant $\omega$, a small level of noise is beneficial. In contrast, the efficiency monotonically decreases with $\epsilon/\omega$ for RTPs. Moreover, the efficiency is consistently greater for RTPs than for AOUPs.

The dependence on the model parameters in $\eta$ is contained in $\Delta w^\engine$, given in Eq.~\eqref{eq:deltawengine}. This term is identical for RTPs and AOUPs and reaches a maximum at a finite persistence time $\tau$, consistent with the findings of Ref.~\cite{malgaretti2022szilard}. 
In the quasistatic limit $\tf \to \infty$, where efficiency is maximized, this optimum $\tau$ is $\tau = 1/k$, and $\Delta w^\engine \to 1/4$---the engine is most efficient when the relaxation timescale in the trap matches the relaxation time of the self-propulsion.
Together, we find an upper bound to the information-efficiency
\begin{align}
    \eta \leq \eta_\mathrm{max} =  \frac{1}{4\ln 2}\, .
\end{align}
The engine is most efficient when run with an RTP for $k\tau=1$, low frequency $1/\tf$, and low error $\epsilon$.

We recall that the extractable work is independent of the model of self-propulsion. From the mathematical structure of the expressions, we expect that this result also holds beyond AOUPs and RTPs (at least in one dimension).
Furthermore, recalling the discrete nature of the $v$ distribution for RTPs, leading to an information aquisition cost of just one bit for small $\epsilon$, we expect RTPs have the smallest information cost of any possible active particle models.

\section{Alternative cost functional incorporating internal dissipation}\label{sec:work-functionals}

Throughout the above, we focused on control that minimizes the external work input required by the controller, which serves as our cost functional in the optimization. The motivation for this choice is twofold: first, this external work input is accessible in experiments, and second, it provides a lower bound to the total energy cost for the ``controller.'' If we view the activity as a given medium that we can exploit, this is the appropriate cost functional. 
At the same time, we may take into account that the particle performs work to actively propel itself, which we here refer to as the internal work. 
Depending on the realization of the active system, control that  minimizes both internal and external work expenditure could be of interest.
In this section, we define 
an alternative cost functional that accounts for the external and internal work. We further examine the optimal open- and closed-loop protocol that solve the optimal control problem in Eq.~\eqref{eq:optimalcontrol} for the total work functional.
Note that this approach still does not explicitly quantify the cost of the underlying mechanism to sustain the dissipation of the active swimmer, which can be included if an explicit model of internal degrees of freedom of the particle is included, see e.g., \cite{gaspard2017communication,pietzonka2017entropy,speck2018active,speck2022efficiency}.

Recall the work required by an external controller in Eq.~\eqref{eq:def_work1}, which we denote in this section $W^\ext $ for clarity, given by
$W^\ext[\lambda(t), x(t)] := \int_0^{\tf}\plaind{t}\ \dot{\lambda} k (\lambda - x)$. 
The work required for self-propulsion to move the particle by a distance $\plaind{x}$ is given by $v \plaind{x}$.
Taking external and internal work together, this leads to the work functional
\begin{equation}
\label{eq:def_work2}
  W^\extint[\lambda(t), x(t)] := \int_0^{\tf}\plaind{t}\ \left( \dot{\lambda} k (\lambda - x) + \dot{x} v \right) \ .
\end{equation} 

Let us now consider the average of this cost functional in the different ensembles. Recalling that $\lambda(t)$ and $\ave{x(t)}_\alpha$ are directly related through Eq.~\eqref{eq:lambda_from_x}, allows us to express the average work functional solely in terms of $\ave{x(t)}_\alpha$, leading to 
\begin{align}
\begin{split}\label{eq:work_extint_functional}
  \ave{ W^\extint }_\alpha
  &= W^\extint \left[  \lambda_\alpha(t), \ave{x(t)}_\alpha \right]
  \\
  &= \frac12k \left. (\lambda - \ave{x}_\alpha)^2 \right\vert_0^{\tf} + \int_0^{\tf}\dt\, \ave{\dot{x}}_\alpha^2
  \\
  & + \int_0^{\tf}\dt\, \left( \Var_\alpha [v(t)] - k \, \Cov_\alpha\, [x(t)v(t)] \right)  \ .
\end{split}  
\end{align}
In contrast to the corresponding expression for $\ave{ W^\ext }_\alpha$ given in Eq.~\eqref{eq:work_ext_functional}, $\ave{ W^\extint }_\alpha$ explicitly features second moments.
Just as in the case of external work Eq.~\eqref{eq:optimaltransport_work_ext}, the control problem can formally be expressed as
\begin{align}\label{eq:optimaltransport_work_extint}
  \argmin_{\ave{x(t)}_\alpha} W^\extint[ \ave{x(t)}_\alpha ] .
\end{align}
Note that as $\ave{ W^\extint }_\alpha$ depends solely on first and second moments of position and self-propulsion, which are identical for RTPs and AOUPs (see Secs.~\ref{sec:averages} and \ref{sec:covariance}), the minimizer $\ave{x(t)}_\alpha$ and the corresponding optimal protocol remain identical for RTPs and AOUPs, just like the protocols optimizing the external work.

\subsection{Optimal open-loop protocols}
For the case of open-loop control, we use the stationary ensemble, where the total average work becomes
\begin{align}\label{eq:opt_work_extint_openloop}
\begin{split}
  \ave{W^\extint}_\op &= \frac{k}2 \left. \left(\lambda - \ave{x}_\op\right)^2 \right\vert_0^{\tf} + \frac{\omega^2}{1 + k \tau} \tf 
  \\
  &\qquad + \int_0^{\tf}\dt\, \ave{\dot{x}}^2_\op  \ .
\end{split}
\end{align}
Here, we used the expressions for the variance and cross-covariance given in Eqs.~\eqref{eq:var_v_stationary} and  \eqref{eq:cov_xv_stationary}.

The second term, $\omega^2\tf/(1 + k\tau)$, accounts for the contribution by the internal work, while the remaining terms are identical to \eqref{eq:work-functional-ext-open-loop}.
Importantly, the second term is independent of $\ave{x}_\op$ and therefore does not contribute towards the optimization. Consequently, the optimal open-loop protocol and $\ave{x}_\op$ are identical for both cost functionals, 
$\ave{W^\ext}_\op$ and $\ave{W^\extint}_\op$, given in Eqs.~\eqref{eq:opt_stationary}.

While the optimal protocol does not change when we include the internal work into the cost functional, the associated cost required to execute the optimal protocol differs between $W^\ext$ and $W^\extint$. 
Evaluating Eq.~\eqref{eq:opt_work_extint_openloop} for the optimal protocol yields 
\begin{equation}\label{eq:opt_work_extin_openloop}
  \ave{W^\extint}_\op = \frac{\lambdaf^2}{\tf + 2/k} + \frac{\omega^2}{1 + k\tau} \tf \ ,
\end{equation}
which cosists of the passive energy found in the open-loop case  (see Eq.~\eqref{eq:opt_work_ext_openloop}) and the additional dissipation term linear in $\tf$ Thus, taking this contribution into account, the work expenditure is strictly larger than for PPs.

Remarkably, the work in Eq.~\eqref{eq:opt_work_extin_openloop} is non-monotonic in $\tf$, with an optimal protocol duration of
\begin{equation}
    \tf^* = \sqrt{ 1+k\tau} \left|\frac{\lambdaf}{\omega}\right|-\frac{2}{k}\,.
\end{equation}
Such an optimal protocol duration has recently been shown to be a typical feature of thermodynamically optimal control of a wide class of active matter systems \cite{davis2024active}.

\subsection{Optimal closed-loop protocols}
\label{sec:opt_protocol_Wextint}

Now we address the closed-loop control with initial measurement of $x$ and $v$, analogously to Sec. \ref{sec:closed-loop}. To this end, we consider Eq.~\eqref{eq:work_extint_functional} in the conditional ensemble, which leads to the cost functional
\begin{equation}\label{eq:Wext_action}
\begin{split}
  &\ave{W^\extint}_{x_0,v_0} = \frac12k \left. (\lambda - \ave{x}_{x_0,v_0})^2 \right\vert_0^{\tf} 
  \\
  &\qquad + \int_0^{\tf}\dt\, \left( \Var_{x_0,v_0} [v(t)] - k \, \Cov_{x_0,v_0}\, [x(t)v(t)] \right) 
  \\
  &\qquad + \int_0^{\tf}\dt\, \ave{\dot{x}}_{x_0,v_0}^2 \ .
\end{split}
\end{equation}
In contrast to the open-loop control, the variance and covariance terms are now time-dependent, see Eqs.~\eqref{eq:var_v} and \eqref{eq:cov_xv}. However, they remain \textit{independent} of the average particle position $\ave{x(t)}_{x_0,v_0}$. As a result, the optimization problem Eq.~\eqref{eq:optimaltransport_work_extint} reduces to that of a passive particle. As a result, the optimal particle position remains identical to the open-loop case and is given by
\begin{subequations}\label{eq:opt_extint_conditional}
\begin{align}\label{eq:opt_x_extint_conditional}
  \ave{x(t)}_{x_0,v_0} = x_0 + \frac{\lambdaf - x_0}{\tf + 2/k} t \ .
\end{align}
The corresponding protocol follows from Eq.~\eqref{eq:lambda_from_x}
\begin{align}
  \lambda_{x_0,v_0}(t) &= \ave{x(t)}_{x_0,v_0} + \frac{\lambdaf - x_0}{k\tf + 2} - \frac{v_0}{k} e^{-t/\tau} \  ,
  \label{eq:opt_lambda_extint_conditional}
\end{align}
\end{subequations}
with boundary conditions $\lambda_{x_0,v_0}(0) = 0$ and $\lambda_{x_0,v_0}(\tf) = \lambdaf$.
In contrast to the position, the protocol is changed compared to the open-loop case due to the $v$-dependence in Eq.~\eqref{eq:lambda_from_x}.
The protocol is nonlinear at intermediate times and has asymmetric jumps at $t=0$ and $t=\tf$ where the initial jump can be in opposite direction of $\lambdaf$, similar to the optimal protocol from optimizing $\ave{W^\ext}_{x_0,v_0}$, see Sec. \ref{sec:closed-loop}.

The corresponding average work per measurement is 
\begin{align}\label{eq:Wextint_conditional}
\begin{split}
 & \ave{W^\extint}_{x_0,v_0} = - \frac12 k x_0^2  + \frac{(\lambdaf - x_0)^2}{\tf + 2/k} + \frac{\omega^2}{1 + k\tau}  \tf
  \\
  &+ \frac{\tau \omega^2}{1 - k\tau}
  \left[ 
    \frac{2k\tau}{(1+k\tau)^2} (1 - e^{-(k + 1/\tau) \tf}) 
    - \frac12  (1 - e^{-2\tf/\tau})
  \right]    
  \ .
\end{split}
\end{align}
Remarkably, the resulting expression is independent of the self-propulsion measurement $v_0$, in contrast to the corresponding expression for $W^\ext$ in Eq.~\eqref{eq:Wextint_conditional}. In the case of $W^\ext$, the orientation of $v_0$ relative to $\lambdaf$ plays a crucial role: it is advantageous when $v_0$ aligns with $\lambdaf$, and disadvantageous when it does not. Moreover, the magnitude of $v_0$ directly influences the extent of this effect.
In contrast, for $W^\extint$, there are no inherently ``good or bad'' measurement outcomes. Regardless of the measurement, the optimal protocol can fully compensate for any additional costs introduced by the self-propulsion via the internal work term in $W^\extint$. 
However, this also implies that self-propulsion measurements do not contribute to work extraction, as is the case for $W^\ext$.

\section{Conclusion}
\label{sec:conclusion}

In this paper, we derived minimum-work protocols for AOUPs and RTPs under open and closed-loop control. 
Since RTPs and AOUPs are identical up to the second moment of the self-propulsion, and the cost function is linear in the particle position, the derived optimal protocols are identical for RTPs and AOUPs. 

In the open-loop case, the optimal protocol and associated work are identical to those for passive particles \cite{schmiedl2007efficiency}. 
This result indicates that, on average, the control cannot capitalize on the activity to improve the protocol without additional information. 
However, the equivalence between active and passive cases is limited to the level of average quantities, and activity changes the work distribution for higher moments, as seen in the work variance which is increased compared to the passive case.
Broadly speaking this extra variance arises because on timescales up to the persistence time, the active particle is either assisting or opposing the external force, albeit with zero mean effect.

For the closed-loop case, we extend the control problem by incorporating measurements of the initial position and self-propulsion. The corresponding optimal protocol deviates from the passive case, with a linear-exponential form and asymmetric jumps.
The work per protocol execution is a balance between a cost term accounting for protocol boundary conditions, and two extraction terms, one extracting initial potential energy and the other extracting active energy. 
Importantly, work is reduced compared to the open-loop case on average due to successful harnessing of activity.
The average work is non-monotonic in the persistence time and reaches a minimum at a finite value. This means that activity in the form of non-equilibrium fluctuations with finite persistence time is beneficial for the closed-loop control, whereas control of infinitely persistent (or passive) particles is more costly on average.

We have focused on protocols that minimize the external work required by the controller to move the trap, which is a quantity that can be measured. We also derived the optimal protocol taking both the external work and the internal dissipation of the active particle into account. The resulting open-loop case resembles that for passive particles with an added dissipation term. The closed-loop case follows a distinct protocol compared to the one that minimizes external work only. Notably, the work in the closed-loop case is independent of the self-propulsion measurement, so that a measurement cannot increase or reduce the total work. 
The dissipation term grows linearly in time, while all other energetic terms decay for large protocol duration. This tradeoff between adiabaticity and dissipation leads to an optimal, finite protocol duration, similar to previous literature results  in which the same tradeoff is observed when minimizing active dissipation~\cite{davis2024active}. 

We further analyzed the effect of measurement uncertainty on the energetics.
Assuming Gaussian measurement error, inexact measurements increase the work quadratically with the error. The protocol remains energetically favorable compared to the open-loop case until the error exceeds the standard deviation of the self-propulsion.
Since our closed-loop protocol assumes exact measurements, it is suboptimal under measurement uncertainty. Optimizing the protocol by taking into account knowledge about the measurement error would improve the protocol, as discussed in earlier studies \cite{abreu2011extracting, cocconi2024efficiency}.


By further optimizing the protocol target position, we derive protocols that extract work for any measurement outcome. Remarkably, this double optimization---first optimizing the work given a target position, and then optimizing the target---yields a protocol qualitatively similar to those obtained from a single, unconstrained optimization, where the force applied to the particle is equal to half of its self-propulsion \cite{malgaretti2022szilard, cocconi2023optimal}.
Applied repeatedly, this protocol allows us to construct a minimal engine which maximizes work extraction from the activity in the absence of further knowledge about the measurement uncertainty, provided that the measurement error remains smaller than the standard deviation of the self-propulsion. 

The fundamental information costs of measurements vary between AOUPs and RTPs. For RTPs, these costs reach a maximum of one bit as the measurement error approaches zero, whereas for AOUPs, the costs are consistently higher costs than for RTPs and diverge in the limit $\epsilon \to 0$.
As a result, the information efficiency, defined by the mean work extraction divided by the information acquisition costs, is highest for RTPs, suggesting that the discreteness of $v$ and associated non-Gaussian fluctuations are beneficial for the engine.
The information efficiency, defined as the ratio of the work extracted divided by the information acquisition cost, reaches a maximum for finite persistence time, and the universal upper efficiency bound for our design of engine is $1/(4\ln 2 )\approx 0.36$. Higher efficiencies can presumably be reached with more complex machines, for instance by dynamically adjusting the trap stiffness.
To find optimal solutions for such processes, one might apply recent machine-learning based algorithms \cite{whitelam2023demon,engel2023optimal,loos2023universal,casert2024learning} to the case of active particles. Moreover, using these methods, more complicated systems like collective active systems could be studied.


To address problems that are not exactly solvable, several approximation schemes have been proposed in the literature. In particular, the versatile approach based on linear response theory \cite{sivak2012thermodynamic,blaber2023optimal} applies to a wide class of problems and has recently been extended to active matter~\cite{davis2024active,gupta2023efficient}. This method is valid in the regime of weak and slow driving, where the particle remains in a stationary state throughout the protocol. In this limit, no driving discontinuities occur. 
Reference \cite{blaber2022optimal} introduces a “strong-trap approximation” for overdamped dynamics, suitable for the opposite regime of strong driving. Additionally, a general framework for vanishingly short protocols $\tf \to 0$ was proposed in~\cite{blaber2021steps}.

In contrast, the results presented in this work are exact.
While our results are consistent with these approximation schemes within their respective regimes ($\tf \to \infty$, $k \to \infty$, $\tf \to 0$), our closed-loop protocol explicitly exploits the regime where the system is not in a steady state at all times, the protocol features finite jumps, and the duration is finite. 
In particular, the third term in Eq.~\eqref{eq:avg_work_partialconditional} utilizes the initial self-propulsion while the system has memory of it. 
Importantly, it is precisely when both $k$ and $\tf$ are finite that net work can be extracted from activity.
Thus, the protocols we derive here constitute a significant addition to the collection of analytically exact results in the study of optimal control in active matter.

\appendix

\subsection*{Acknowledgements}
We thank Robert Jack and \'Edgar Rold\'an for insightful discussions.
R.G.-M. acknowledges support from a St John's College Research Fellowship, Cambridge.
J.S. acknowledges funding through the UK Engineering and Physical Sciences Research Council (Grant number 2602536). 
S.L. acknowledges funding through the postdoctoral fellowship of the Marie Skłodowska-Curie Actions (Grant Ref. EP/X031926/1) undertaken by the UKRI, through the Walter Benjamin Fellowship (Project No. 498288081) from the 
Deutsche Forschungsgemeinschaft (DFG), and from Corpus Christi College, Cambridge.

\section{Covariance of position and self-propulsion}
\label{sec:covariance}

In this subsection, we derive the covariance of position and self-propulsion, extending the discussion of Sec.~\ref{sec:averages}.
We consider the centralized fluctuations of the position and self-propulsion in the ensemble $\alpha$ as
\begin{subequations}
\begin{align}
  \Delta v(t) &:= v(t) - \ave{v(t)}_\alpha \ ,
  \\
  \Delta x(t) &:= x(t) - \ave{x(t)}_\alpha \ .
\end{align}
\end{subequations}
These fluctuations are related to the covariance and variance through
\begin{subequations}
\begin{align}
\begin{split} \label{eq:var_v_defn}
  \Var_\alpha [v(t)] &:= \ave{v(t)^2}_\alpha - \ave{v(t)}^2_\alpha 
  \\
  &=  \ave{\Delta v(t)^2}_\alpha \ ,
\end{split}
\\
\begin{split}
  \Cov_\alpha[v(t) v(t')] &:= \ave{v(t)v(t')}_{\alpha} - \ave{v(t)} \ave{v(t')}_\alpha 
  \\
  &=  \ave{\Delta v(t) \Delta v(t')}_\alpha \ ,
\end{split}
\end{align}
\end{subequations}
with analogous expressions for $x(t)$.
For an RTP, the propagator in $v$ from time $t_0$ to $t$ is given by
\begin{subequations}\label{eq:v_formal_solution_RTP}
\begin{align}
  P\big(v(t) = +\omega | v(t_0) = \pm \omega\big) &= \frac12 \left(1 \pm e^{-(t-t_0)/\tau}\right) \ ,
  \\
  P\big(v(t) = -\omega | v(t_0) = \pm \omega\big) &= \frac12 \left(1 \mp e^{-(t-t_0)/\tau}\right) \ ,
\end{align}
\end{subequations} 
which leads to the self-propulsion correlations
\begin{align}
\begin{split}
  \ave{ v(t) v(t') }_\alpha^\text{RTP}  
  &= \omega^2 e^{- |t-t'| / \tau} \ ,
\end{split}
\end{align}
which are identical across all ensembles $\alpha$.
This implies the following covariance of $v(t)$ for an RTP,
\begin{equation}
\begin{split}\label{eq:Sv_corr_RTP}
  \Cov_\alpha[v(t) v(t')]^\text{RTP} &= \omega^2 \left( e^{- |t-t'|/\tau} - e^{-(t+t'-2t_0^\alpha)/\tau} \right) \ ,
\end{split}
\end{equation}
where $t_0^\alpha \to -\infty$ in the stationary ensemble and $t_0^\alpha = 0$ in the conditional and partial conditional ensemble. 

For AOUPs, the propagator in $v$ is given by
\begin{align}\label{eq:aoup_v_propagator}
\begin{split}
  P(&v(t) = v | v(t_0) = v_0) \\
  &= \mathcal{N}\left(v_0 e^{-(t-t_0)/\tau}, \omega^2(1-e^{-2(t-t_0)/\tau}) \right) \ ,
\end{split}
\end{align}
where $\mathcal{N}(\mu,\sigma^2)$ denotes a Gaussian with mean $\mu$ and variance $\sigma^2$. 
Integrating Eq.~\eqref{eq:v_AOUP} gives
\begin{align}
  \Delta v(t) &= \frac{2D_v}{\tau^2} \int_{t_0^\alpha}^t \dt'\, e^{(t'-t)/\tau} \xi (t') \ .
\end{align}
Using $\ave{ \xi(t) \xi(t’) } = \delta(t-t’)$, the covariance for AOUPs becomes
\begin{equation}\label{eq:cov_v_aoup}
  \Cov_{\alpha}[v(t) v(t')]^\text{AOUP} = \frac{D_v}{\tau} \left(e^{-|t-t'|/\tau} - e^{-2(t + t' - 2t_0^\alpha)/\tau}\right) \ .
\end{equation}
Comparing Eqs.~\eqref{eq:cov_v_aoup} and \eqref{eq:Sv_corr_RTP} shows that the covariances are equivalent when $D_v/\tau = \omega^2$, confirming the parameter matching  in Eq.~\eqref{eq:rtp_aoup_match}. 
In the following, we adopt the convention of using $\omega^2$ instead of $D_v/\tau$ to unify notation. 
The covariance of $v$ obeys
\begin{subequations}\label{eq:cov_v}
\begin{align}
  \Cov_\op[v(t) v(t')] &= \omega^2 e^{- |t-t'|/\tau} \ ,
  \\
  \Cov_{x_0,v_0}[v(t) v(t')] &= \Cov_{v_0}[v(t) v(t')]
  \\
  \nonumber
  &= \omega^2 \left( e^{- |t-t'|/\tau} - e^{-(t+t')/\tau} \right) \ ,
\end{align}
\end{subequations}
and the variance obeys
\begin{subequations}\label{eq:var_v}
\begin{align}
\label{eq:var_v_stationary}
  \Var_\op [v(t)] &= \omega^2 \ ,
  \\
\begin{split}
\label{eq:var_v_conditional}
  \Var_{x_0,v_0} [v(t)] &= \Var_{v_0} [v(t)] 
  \\
  &= \omega^2\left(1 - e^{-2t/\tau}\right) \ .
\end{split}
\end{align}
\end{subequations}

Next, we analyze the covariance of the position. Integrating Eq.~\eqref{eq:oLE}, the excess position is given by
\begin{equation}
\begin{split}
  \Delta x_{\alpha}(t) &= \int_{t_0^\alpha}^t \dt'\, e^{k(t'-t)}  \Delta v_{\alpha}(t') 
  \\
  &+ \sqrt{2D} \int_{t_0^\alpha}^t \dt'\, e^{k(t'-t)}  \xi(t')
  \ .
\end{split}
\end{equation}
Since $\Delta_\alpha v$ is independent of $\xi$, and $\xi$ is $\delta$-correlated, we can express the positional covariance in terms of the covariance of $v$,
\begin{align}
\nonumber
  &\Cov_{\alpha}[x(t)x(t')] = \ave{ \Delta x_\alpha(t) \Delta x_\alpha(t') }_{\alpha} 
  \\
  &= \frac{D}{k} \left(e^{-k|t-t'|} - e^{-k(t+t' - 2t_0^\alpha)}\right) 
  \\
  \nonumber
  &\quad + \int_{t_0^\alpha}^t \dt_1\int_{t_0^\alpha}^{t'} \dt_2\, e^{k(t_1-t)} e^{k(t_2-t')}  \ave{ \Delta v(t_1) \Delta v(t_2) } \ .
\end{align}
Here, the first term arises from the correlations in the noise $\xi$. Evaluating the integral, the positional covariance obeys
\begin{align}
\label{eq:Covx_1}
  \Cov_\op[x(t)&x(t')] = \frac{D}{k}  e^{-k|t-t'|}  
  \\
  \nonumber
  &+ \frac{\omega^2 }{k^2 - 1/\tau^2}\left( e^{-|t-t'|/\tau} - \frac1{k\tau} e^{-k|t-t'|}\right)  \ .
\end{align} 
The positional variance in the stationary case is given by
\begin{equation}\label{eq:var_x_stationary}
  \Var_\op [x(t)] = \frac1k \left(D + \frac{\omega^2}{k + 1/\tau}\right)  \ .
\end{equation}

Finally, we calculate the cross-correlations between $x(t)$ and $v(t)$
\begin{align}
\nonumber
  \Cov_{\alpha}[x(t) v(t)] &= \ave{x(t) v(t)}_{\alpha} - \ave{x(t)}_{\alpha} \ave{v(t)}_{\alpha}
  \\
  \label{eq:cov_xv_defn}
  &= \ave{\Delta x(t) \Delta v(t)}_\alpha
  \\
  \nonumber
  &= \int_{t_0^\alpha}^t \dt_1\, e^{k(t_1-t)}   \ave{ \Delta v(t_1) \Delta v(t) }_{\alpha}
 \ , 
\end{align}
which using \eqref{eq:cov_v} leads to
\begin{subequations}\label{eq:cov_xv}
\begin{align}
\label{eq:cov_xv_stationary}
  \Cov_\op[x(t)v(t)] &= \frac{\omega^2}{k+1/\tau}  \ ,
  \\
\begin{split}
\label{eq:cov_xv_conditional}
  \Cov_{x_0,v_0}[x(t)v(t)] &= \frac{\omega^2}{k+1/\tau} \left(1 - e^{-(k+1/\tau)t}\right) 
  \\
  &+ \frac{\omega^2}{k-1/\tau} e^{-t/\tau} (e^{-kt} - e^{-t/\tau}) \ .
\end{split}
\end{align}
\end{subequations}

Note that the variances of $x$ and $v$, as well as cross-covariance in the stationary ensemble, given by Eqs.~\eqref{eq:var_v_stationary}, \eqref{eq:var_x_stationary} and \eqref{eq:cov_xv_stationary}, are time-independent and match the stationary expressions in Eq.~\eqref{eq:steadystate}, as expected.

\section{The impact of measurement uncertainty}
\label{app:measurement_uncertainty}

In this section, we provide additional technical background for the calculation measurement uncertainty discussed in Sec.~\ref{sec:epsilon}. 

\subsection{Probability density function of the true and the measured system state}\label{app:PDFs}

In the main text, we introduced the quantities $\vgt$ and $\vm$ to denote the true and measured values of $v(0)$, respectively. 
We assumed that the measured value $\vm$ given the true value $\vgt$ is normal distributed around the true value with error $\epsilon^2$
\begin{equation}\label{eq:measurementassump}
  P(\vm|\vgt)=\mathcal{N}_{\vm}(\vgt,\epsilon^2) \ .
\end{equation} 
Here and in the following, we use an index to specify the random variable of a normal density, i.e., 
\begin{equation}\label{eq:gaussian_notation}
  \mathcal{N}_{X}(Y,\epsilon^2) = \frac1{\sqrt{2\pi} \epsilon} e^{{-({X-Y})^2}/(2\epsilon^2)} \ .
\end{equation}
This allowed us to introduce the corresponding ensemble $\ave{\bullet}_{\vgt,\vm}$. 
The total average over initial condition is now a dual average over $\vgt$ and $\vm$ 
\begin{equation}
  \E_{\vgt,\vm}[\bullet] := \int\plaind{\vm}\plaind{\vgt} \ \bullet P(\vm, \vgt) \ .
\end{equation}
We calculate the joint density $P(\vm, \vgt)$ for both AOUPs and RTPs in the following. 

The true values $\vgt$ are distributed according to the (pre-measurement) steady-state densities, $P(\vgt)$, which are different for AOUPs and RTPs.
For AOUPs, the self-propulsion is normal-distributed, $P(\vgt)=\mathcal{N}_{\vgt}(0,{\omega}^2)$, with $\omega^2 = D_v/\tau$, where we make use of the notation $\mathcal{N}_{X}(Y,\epsilon^2)$ in Eq.~\eqref{eq:gaussian_notation} to specify the density of a normal random variable $X$. 
Accordingly, the joint probability density of the self-propulsion and its measurement outcome is 
\begin{equation}
\begin{split}
\elabel{joint-AOUP}
  P(\vgt,\vm) &= P(\vgt)P(\vm|\vgt) 
  \\
  &= \mathcal{N}_{\vm}(\vgt,\epsilon^2) \mathcal{N}_{\vgt}(0,{\omega}^2) \ . 
\end{split}
\end{equation}
Marginalising over $\vgt$, the probability density of the measurement outcome is 
\begin{equation}
\elabel{marginal-AOUP}
    P(\vm) =\int_\mathbb{R} \mathrm{d}\vgt\, P(\vgt,\vm) =\mathcal{N}_{\vm}(0,{\omega}^2+\epsilon^2) \ .
\end{equation}

For RTPs, the steady-state probability density of the self-propulsion is given by 
\begin{equation}
    P(\vgt)= \frac12 [\delta(\vgt+{\omega})+\delta(\vgt-{\omega})] \ .
\end{equation}
Following the same steps as for AOUPs, the joint probability density for RTPs is
\begin{align}
\begin{split}
\label{eq:joint-RTPs}
  P(\vgt,\vm) &= P(\vgt)P(\vm|\vgt) 
  \\
  &= \frac12\mathcal{N}_{\vm}(\vgt,\epsilon^2)  [\delta(\vgt+{\omega})+\delta(\vgt-{\omega})] \ ,
\end{split}
\end{align}
and the marginalized probability density of the measurement outcome is
\begin{equation}\label{eq:marginal-RTP}
\begin{split}
  P(\vm) &= \int_\mathbb{R} \mathrm{d}\vgt \, P(\vgt,\vm)  
  \\
  &=
  \frac12 \left[\mathcal{N}_{\vm}(-{\omega},\epsilon^2) + \mathcal{N}_{\vm}({\omega},\epsilon^2)\right] \ .
\end{split}
\end{equation}

\subsection{Calculation of the additional work input due to the presence of measurement uncertainty}
\label{app:Work-measurement-uncertainty}

\begin{table*}[hbt]
\def\arraystretch{1.5}
\setlength\tabcolsep{7pt}
\begin{tabular}{l|cccc}
& true value & measurement value & difference & average difference \\
\hline
Measurement & $\vgt$ & $\vm$ & $\vgt - \vm$ & 0 
\\
Avg. position & $\ave{x(t)}_{\vgt,\vm}^{\rm r} := \ave{x(t)}_{\vgt}[\lambdam] $ & $\ave{x(t)}_{\vm}^{\rm m} := \ave{x(t)}_{\vm}[\lambdam] $ & $\Delta x_{\vgt,\vm}(t)$ & 0 
\\
Avg. work & $\ave{W}_{\vgt,\vm}^{\rm r} := W\left[\lambdam, \ave{x(t)}_{\vgt,\vm}^{\rm r}\right]$ & $\ave{W}_{\vm}^{\rm m} := W\left[\lambdam, \ave{x(t)}_{\vgt,\vm}^{\rm m}\right]$ & $\Delta W_{\vgt,\vm}$ & $2\tau \epsilon^2 \Delta w$
\end{tabular}
\caption{Overview over quantities introduced in App.~\ref{app:Work-measurement-uncertainty}.}
\end{table*}

Executing the protocol $\lambdam$ from Eq.\eqref{eq:lambdam} incurs additional energetic costs compared to the optimal protocol in Eq.\eqref{eq:opt_partialconditional}, as a direct consequence of measurement uncertainty. The explicit expression for this increased average work was given in Eq.~\eqref{eq:avgavg_work_w_uncertainty} in the main text. In this section, we present the detailed derivation of that result.

Note that we can write the protocol $\lambdam$ in Eq.~\eqref{eq:lambdam} as
\begin{align}
    \lambdam(t) = \ave{x(t)}_{\vm}^{\rm m} + \frac{d_{\vm}^{\rm m}}{k\tf + 2} - \frac{\vm}{2 k} e^{-t/\tau}
  \ ,
\end{align}
with $\lambdam(0) = 0, \lambdam(\tf) = \lambdaf$ and
where $\ave{x(t)}_{\vm}^{\rm m}$ and $d_{\vm}^{\rm m}$ are obtained from $\ave{x(t)}_{v_0}$ and $d_{v_0}$ by substituting $v_0$ by $\vm$, the latter defined in Eq.~\eqref{eq:d_vm^m} while the former is explicitly given by
\begin{align}
  \ave{x(t)}_{\vm}^{\rm m} &:= \ave{x(t)}_{\vm}[\lambdam] 
  \\
  \nonumber
  &\;=  \frac{\vm}{k+1/\tau}
  + \frac{d^{\rm m}_{\vm}}{\tf + 2/k} t  + \frac{\tau \vm}{2}  (1-e^{-t/\tau}) \ .
\end{align}
Here (and in the main text), we use the superscript m to denote quantities which are obtained by naively substituting $v_0$ by $\vm$ in quantities derived without measurement uncertainty (e.g. $d_{\vm}^{\rm m}$ derives from $d_{\vm}$ by substituting $v_0$ by $\vm$). 
Crucially, however, $\ave{x(t)}_{\vm}^{\rm m}$ does not represent the mean particle position, unless $\vm$ were the true value of $v(0)$. 
Instead, we obtain the true average particle position $\ave{x(t)}_{\vgt,\vm}^{\rm r}$ from the general expression in Eq.~\eqref{eq:ave_x_formal1}, by substituting $v_0$ with $\vgt$ and using the protocol $\lambdam$ as given in Eq.~\eqref{eq:lambdam} for $\lambda$, i.e., 
\begin{align}
\label{eq:avg_x_uncertain_measurement1}
  \ave{x(t)}_{\vgt,\vm}^{\rm r} &:= \ave{x(t)}_{\vgt}[\lambdam] 
  \\
  \nonumber
  &\;=\frac{ \vgt}{k+1/\tau} e^{-k t}
  + \frac{\vgt}{k-1/\tau} \left(
  e^{-t/\tau} - e^{-k t}
  \right) 
  \\
  \nonumber
  &\;\quad + k \int_{0}^t\dint{t'} \lambdam(t')e^{-k(t-t')} \ .
\end{align}
Here and in the following, we use the superscript r to distinguish quantities which account for the measurement uncertainty from the ones with superscript m. 
Notice that $\ave{x(t)}_{\vgt,\vm}^{\rm r}$ depends on both the true value ($\vgt$, first two terms) and the measurement value ($\vm$, through $\lambdam$ in the final term). 
We define the difference between $\ave{x(t)}_{\vgt,\vm}^{\rm r}$ and $\ave{x(t)}_{\vm}^{\rm m}$ as
\begin{align}\label{eq:sumX}
  \Delta x_{\vgt,\vm}(t) &:= \ave{x(t)}_{\vgt,\vm}^{\rm r} - \ave{x(t)}_{\vgt,\vm}^{\rm m} 
  \\
  \nonumber
  &\;= \frac{\vgt - \vm}{k + 1/\tau} e^{-kt} + \frac{\vgt - \vm}{k-1/\tau}\left(e^{-t/\tau} - e^{-k t}\right) \ ,
\end{align}
where the second equality follows from Eq.~\eqref{eq:ave_x_formal1}.
In the absence of a measurement error, we have that $\vgt=\vm$ and $\Delta x_{\vgt,\vm}$ vanishes. 

The total true average work accounting for the measurement uncertainty was defined in Eq.~\eqref{eq:Wtotal_SM}, obtained by evaluating the work functional in Eq.~\eqref{eq:workfunctional} using $\lambdam$ and $\ave{x(t)}_{\vgt,\vm}^{\rm r}$. We here recall the definition for convenience
\begin{equation}
\begin{split}
  \ave{ W }_{\vgt,\vm}^{\rm r} &:= W[\lambdam(t), \ave{x(t)}_{\vgt,\vm}^{\rm r}] 
  \\
  &= k \int_0^{\tf}\plaind{t}\ \dot{\lambda}_{\text{m}} \left(\lambdam - \ave{x}_{\vgt,\vm}^{\rm r} \right) \ .
\end{split}
\end{equation}
Using \Eref{sumX}, we can split Eq.~\eqref{eq:Wtotal_SM} into two contributions
\begin{align}
\nonumber
    \ave{ W }_{\vgt,\vm}^{\rm r} &= \ k \int_0^{\tf}\plaind{t}\ \dot{\lambda}_{\text{m}}  \left(\lambdam - \ave{x}_{\vm}^{\rm m} - \Delta x_{\vgt,\vm}\right) \\
\label{eq:Wtot_defcost}
    &= \ave{ W }_{\vm}^{\rm m}  + \Delta W_{\vgt,\vm} \ ,
\end{align}
where $\ave{ W }_{\vm}^{\rm m} := W\left[\lambdam(t), \ave{x(t)}_{\vm}^{\rm m}\right]$ 
is the work by the protocol $\lambdam$ if there was no measurement uncertainty [given by substituting $v_0 \to \vm$ in Eq.~\eqref{eq:avg_work_partialconditional}], and $\Delta W_{\vgt,\vm}$ accounts for the additional energetic cost due to the measurement uncertainty.

Using the decomposition in Eq.~\eqref{eq:Wtot_defcost}, the average of $\ave{W}^{\rm  r}_{\vgt,\vm}$ over $\vgt$ and $\vm$ is given by
\begin{equation}
\begin{split}\label{eq:dnsakjdnsad}
  \E_{\vgt, \vm} [ \ave{W}_{\vgt,\vm}^{\rm r} ] 
  &= \E_{\vgt,\vm}[ \ave{W}_{\vgt,\vm}^{\rm m} ] 
  \\
  &\, + \E_{\vgt,\vm}[ \Delta W_{\vgt,\vm} ] \ .
  \end{split}
\end{equation}
To evaluate the first average on the right-hand side of Eq.~\eqref{eq:dnsakjdnsad},
notice that using the Gaussian assumption on $P(\vm|\vgt)$ in Eq.~\eqref{eq:measurementassump} we find 
\begin{align}
  \E_{\vgt,\vm}[(\vm)^2] = \E_{\vgt}[(\vgt)^2 + \epsilon^2] = \omega^2 + \epsilon^2 \ .
\end{align}
As a result, we can obtain $\E_{\vgt,\vm}[ \ave{W}_{\vgt,\vm}^{\rm m} ]$ from $\E_{v_0}[ \ave{W}_{v_0} ]$ given in Eq.~\eqref{eq:avgavg_work_decomposition} by substituting $\omega^2 \to \omega^2 + \epsilon^2$, i.e.,
\begin{equation}\label{eq:sndasjkdnsa}
  \E_{\vgt,\vm}[ \ave{W}_{\vgt,\vm}^{\rm m} ] = \ave{W}_\op - (\omega^2+\epsilon^2) \tau \Delta w \ .
\end{equation}

We next evaluate the second average on the right-hand side of Eq.~\eqref{eq:dnsakjdnsad}.
Starting from \Eref{Wtot_defcost}, we can rewrite the additional cost term $\Delta W_{\vgt,\vm} $ using partial integration
\begin{align}\elabel{integralDelW}
  \Delta W_{\vgt,\vm} &= - k \int_0^{\tf}\plaind{t}\ \dot{\lambda}_{\text{m}}(t) \Delta x_{\vgt,\vm} 
  \\
  \nonumber
  &= - k {\lambdaf} \Delta x_{\vgt,\vm}(\tf) + k \int_0^{\tf}\plaind{t}\ {\lambda}_{\text{m}} \Delta \dot{x}_{\vgt,\vm} \ .
\end{align}
To evaluate the average $\E_{\vgt,\vm}[\Delta W_{\vgt,\vm}]$ over $\vgt$ and $\vm$ with joint probability distribution given in \eref{joint-AOUP} for AOUPs and \eref{joint-RTPs} for RTPs, we only require the following two results
\begin{subequations}\elabel{integralsv}
\begin{align}
    \E_{\vgt,\vm}[\vgt - \vm] &= 0  \ ,
    \\ 
    \E_{\vgt,\vm}[(\vgt - \vm)\vm] &= -\epsilon^2 \ ,
\end{align}
\end{subequations}
because of the linearity of the average.
Using the expressions for $\lambdam$ and $\Delta x_{\vgt,\vm}$ in Eqs.~\eqref{eq:lambdam} and \eqref{eq:sumX} into \eqref{eq:integralDelW} together with Eqs. \eqref{eq:integralsv}, we obtain the average additional energetic cost due to the measurement uncertainty,
\begin{equation}\label{eq:avg_correction}
  \E_{\vgt,\vm}[\Delta W_{\vgt,\vm}] = 2 \tau \epsilon^2 \Delta w \ .
\end{equation}
Recall that $\Delta w$, defined in Eq.~\eqref{eq:norm_excess_work}, is the non-dimensionalized average excess work compared to the open-loop work $\ave{W}_\op$.
Because $\Delta w$ is non-negative, see Eq.~\eqref{eq:pce_nonpositive}, the additional energetic cost is also non-negative,
\begin{equation}
  \E_{\vgt,\vm}[ \Delta W_{\vgt,\vm} ] \geq 0 \ .
\end{equation}
This is expected, as $\lambdam$ is a protocol which is no longer optimal in the presence of measurement uncertainty.

Combining the results in Eqs.~\eqref{eq:sndasjkdnsa} and \eqref{eq:avg_correction} into \eqref{eq:dnsakjdnsad}, we obtain the final result of the average work under measurement uncertainty
\begin{align}
\begin{split}
  \E_{\vgt,\vm}[ \ave{W}_{\vgt,\vm}^{\rm r} ] &= \ave{W}_\op - (\omega^2 - \epsilon^2) \tau \Delta w \ ,
\end{split}
\end{align}
which is the result stated in Eq.~\eqref{eq:avgavg_work_w_uncertainty} in the main text. Note the change of sign inside the round brackets compared to Eq.~\eqref{eq:sndasjkdnsa}.

\section{Engine protocol in the presence of correlations} 
\label{app:engine_mc}

In this appendix, we explain how to generalize the engine protocol $\lambda_{v_0}^\engine$ in Eq.~\eqref{eq:opt_engine}, which relies on steady-state measurements of $v$, to measurements $v_n$ for $n>0$ which are not fully relaxed to the steady-state and remain correlated to the preceding measurement $v_{n-1}$. 
These correlations are relevant during a repeated execution of the engine protocol and require adaptation of the protocol for subsequent cycles $n>0$. 
In the main text, we focus on the case of long cycle periods, 
$\tf/\tau \gg 1$, where the measurements $v_0,v_1, v_2,\ldots$ are approximately uncorrelated. Here, we provide the additional background to address situations where the correlations cannot be neglected.

As the self-propulsion is independent of the particle and trap positions, and the $v$ dynamics is Markovian, a self-propulsion measurement only depends on the previous measurement. As a result, for cycles $n>0$ a measurement $v_n$ is drawn from a partially relaxed state conditioned on the outcome of the previous measurement, and the sequence of measurements $\mathbf{v}_n = (v_0,v_1,v_2,\ldots,v_{n-1})$ forms a Markov chain. 
The transition densities for RTPs is given by Eq.~\eqref{eq:v_formal_solution_RTP} by
\begin{subequations}
\begin{align}
\begin{split}
  P_{\rm RTP}(v_n | v_{n-1}) &= \frac12(1+e^{-\tf/\tau}) \delta(v_n - v_{n-1}) 
  \\
  &\, + \frac12(1-e^{-\tf/\tau}) \delta(v_n + v_{n-1})
  \end{split}
\end{align}
and for AOUPs via Eq.~\eqref{eq:aoup_v_propagator} by
\begin{equation}
  P_{\rm AOUP}(v_n | v_{n-1}) = \mathcal{N}_{v_n}\left(v_{n-1}e^{-\tf/\tau}, \omega^2 (1-e^{-2\tf/\tau}) \right)
\end{equation}
\end{subequations}
The Markov chain is initialized at stationarity.

We define the conditional expectation as
\begin{equation}
  \E_{v_n}[\bullet | v_{n-1}] = \int \plaind{v_n}\, \bullet \, P(v_n | v_{n-1})  \ ,
\end{equation}
and the trajectory expectation
\begin{equation}
  \E_{\mathbf{v}_n}[\bullet] = \int \plaind{v_n}\, \bullet \, P(\mathbf{v}_n) \ ,
\end{equation}
with
\begin{equation}
  P(\mathbf{v}_n) = P(v_n | v_{n-1}) P(v_{n-1} | v_{n-2}) \ldots P(v_1 | v_{0}) P(v_0) \ .
\end{equation}
For both RTP and AOUP, the conditionally average self-propulsion is given by
\begin{equation}
  \E_{v_n}[v(t) | v_{n-1}] = v_{n-1} e^{-t/\tau} \ .
\end{equation}

Because the self-propulsion cannot fully relax between measurements, each  measurement $v_n$ for $n>0$ contains less information about the average particle position at the start of a cycle compared to a measurement at steady conditions. 
Recall that under steady conditions, the average particle position given a measurement is given by Eq.~\eqref{eq:x_given_v_ss}. Now, for partially relaxed measurements, it is given by
\begin{align}\label{eq:x_given_vn}
\begin{split}
  x_{v_n} &:= \ave{x(0)}_{v_n}^\engine 
  \\
  &\;=  
  \begin{cases}
    \frac{1}{k + 1/\tau} v_0 & n = 0
    \\
    \Psi \frac{1}{k + 1/\tau} (v_n - v_{n-1}e^{-\tf/\tau}) & n > 0
  \end{cases}
   \ ,
\end{split}
\end{align}
where the dimensionless function $\Psi = \Psi(\tf,k,\tau)$ is given by
\begin{align}
  &\Psi := (k + 1/\tau) \frac{ \Cov_{x_0,v_0} x(\tf) v(\tf) }{\Var_{x_0,v_0}(\tf)}
  \\
  &\;= \left(1 + \frac{(1 + k\tau) e^{-\frac{2\tf}{\tau}} - 2e^{-\left(k+\frac{1}{\tau}\right)\tf}}{1 - k^2\tau^2}\right) \frac{1 + \coth(\frac{\tf}{\tau})}{2} \ .
\end{align}
The function $\Psi$ is monotonic in $\tf$, and in the limit of full relaxation, $\tf/\tau \to \infty$, $\Psi$ approaches $1$, while it vanishes for vanishing cycle duration, $\tf/\tau \to 0$.

Together, we find the general engine protocol from Eq.~\eqref{eq:opt_engine} by substituting $\xvo \to x_{v_n}$
\begin{subequations}
\begin{align}
  \ave{x(t)}^\engine_{v_n} &= x_{v_n}  + \frac{\tau v_n}{2}  (1-e^{-t/\tau})  \ ,
  \\
  \lambda^\engine_{v_n}(t) &=
  \begin{cases}
      0 & t=0 \\
      \ave{x(t)}_{v_n}^\engine - \frac{v_n}{2 k} e^{-t/\tau} & 0 < t < \tf \\
      x_{v_n} + \frac{\tau v_n}2  (1 - e^{-\tf / \tau}) & t =\tf
  \end{cases}
  \ ,
\end{align}
\end{subequations}
This protocol $\lambda^\engine_{v_n}(t)$ is valid for all values of $\tf$ and generalizes the protocol $\lambda_{v_0}^\engine$ in Eq.~\eqref{eq:opt_engine} of the main text. 

The associated average work per protocol execution is
\begin{equation}
\begin{split}
  \ave{W}_{v_n}^\text{engine} &=  
    - \frac12 k x_{v_n}^2
    - \frac{\tau v_n^2}8  \left(1- e^{-2\tf/\tau} \right) \ .
\end{split}
\end{equation}
To calculate the conditional average work $\E_{v_n}[\ave{W}_{v_n}^\text{engine} | v_{n-1}]$, we first calculate the average square initial position for $n>0$ using Eq.~\eqref{eq:x_given_vn}
\begin{equation}
  \E_{\mathbf{v}_n}[ x_{v_n}^2 ] = \Psi^2 \frac{\tau^2 \omega^2}{(1 + k\tau)^2} (1-e^{-2\tf/\tau})
\end{equation}
and similarly for the average square measurement 
\begin{align}
\begin{split}
  \E_{\mathbf{v}_n}[ v_n^2 ] &= \omega^2(1-e^{-2\tf/\tau}) + \E_{\mathbf{v}_{n-1}}[ v_{n-1}^2] e^{-2\tf/\tau}
  \\
  &= \omega^2 \ ,
\end{split}
\end{align}
where the second equality follows from solving the recursion relation. 
Together, the average work becomes
\begin{align}
\nonumber
  \E_{\mathbf{v}_n}\!\left[ \ave{W}_{v_n}^\text{engine} \right] &=  
    - \frac12 \frac{k \omega^2}{(k + 1/\tau)^2} \Psi^2 \left(1- e^{-2\tf/\tau} \right)
    \\
    &\quad - \frac{\tau \omega^2}8  \left(1- e^{-2\tf/\tau} \right)
    \ .
\end{align}
This average work $\E_{\mathbf{v}_n}\! [ \ave{W}_{v_n}^\text{engine} ]$ generalizes the expression in Eq.~\eqref{eq:avg_work_engine} of the main text. 
The factor $\Psi^2 \left(1- e^{-2\tf/\tau} \right)$ is new, \textit{c.f.} Eq.~\eqref{eq:avg_work_engine}.

\bibliography{bib.bib}

\end{document}